\documentclass[12pt]{article}
\usepackage[utf8]{inputenc}
\usepackage{geometry}
\usepackage{comment}
\usepackage{bbm}
\usepackage{amssymb,amsfonts,amsmath}
\usepackage{float}
\usepackage{hyperref}
\usepackage{enumitem}
\usepackage{graphicx, xcolor, tcolorbox} 
\usepackage{caption, subcaption}
\usepackage[linesnumbered,ruled,vlined]{algorithm2e}
\SetKwInput{KwGiven}{Given}
\usepackage{url}
\usepackage{natbib}
\RestyleAlgo{ruled}
\makeatletter
\renewcommand{\AlCapFnt}[1]{\normalfont\mdseries#1}
\renewcommand{\AlTitleFnt}[1]{\normalfont\mdseries\emph{#1}\unskip}
\renewcommand{\algocf@captiontext}[2]{%
  \AlCapFnt{#1\algocf@typo. }\AlTitleFnt{#2}%
}
\def\@algocf@capt@plain{top}
\renewcommand{\algocf@makecaption}[2]{%
  \addtolength{\hsize}{\algomargin}%
  \sbox\@tempboxa{\algocf@captiontext{#1}{#2}}%
  \ifdim\wd\@tempboxa >\hsize
    \hskip .5\algomargin%
    \parbox[t]{\hsize}{\algocf@captiontext{#1}{#2}}%
  \else
    \global\@minipagefalse%
    \hbox to\hsize{\box\@tempboxa}%
  \fi
  \addtolength{\hsize}{-\algomargin}%
}
\makeatother

\SetKwComment{Comment}{/* }{ */}

\usepackage{authblk}
\newcommand*\samethanks[1][\value{footnote}]{\footnotemark[#1]}

\title{Conditioning on posterior samples for flexible frequentist goodness-of-fit testing}
\author[1]{Ritwik Bhaduri\thanks{These authors contributed equally.}}
\author[2]{Aabesh Bhattacharyya\samethanks}
\author[2]{Rina Foygel Barber}
\author[1]{Lucas Janson}

\affil[1]{Department of Statistics, Harvard University}
\affil[2]{Department of Statistics, University of Chicago}
\date{}
\usepackage{amsmath, amssymb, mathrsfs}

\makeatletter

\newcommand*{\centernot}{%
  \mathpalette\@centernot
}
\def\@centernot#1#2{%
  \mathrel{%
    \rlap{%
      \settowidth\dimen@{$\m@th#1{#2}$}%
      \kern.5\dimen@
      \settowidth\dimen@{$\m@th#1=$}%
      \kern-.5\dimen@
      $\m@th#1\not$%
    }%
    {#2}%
  }%
}

\makeatother

\usepackage{xparse}
\usepackage{amsthm}

\newtheorem{Theorem}{Theorem}[section]

\newtheorem{Definition}{Definition}[section]

\newtheorem{proposition}{Proposition}[section]
\newtheorem{Lemma}{Lemma}[section]

\usepackage{cleveref}
\Crefformat{Theorem}{#2Theorem #1#3}
\Crefformat{proposition}{#2Proposition #1#3}
\Crefformat{corollary}{#2Corollary #1#3}
\Crefformat{claim}{#2Claim #1#3}
\Crefformat{Lemma}{#2Lemma #1#3}
\Crefformat{Definition}{#2Definition #1#3}
\Crefformat{assumption}{#2Assumption #1#3}
\crefformat{assumption}{#2assumption #1#3}

\usepackage{listings}
\begin{document}
\maketitle

\begingroup
\renewcommand\thefootnote{\fnsymbol{footnote}}
\maketitle
\endgroup

\begin{abstract}
Tests of \emph{goodness of fit} are used in nearly every domain where statistics is applied. One powerful and flexible approach is to sample artificial data sets that are exchangeable with the real data under the null hypothesis (but not under the alternative), as this allows the analyst to conduct a valid test using \emph{any} test statistic they desire. Such sampling is typically done by conditioning on either an exact or approximate sufficient statistic, but existing methods for doing so have significant limitations, which either preclude their use or substantially reduce their power or computational tractability for many important models. In this paper, we propose to condition on samples from a Bayesian posterior distribution, which constitute a very different type of approximate sufficient statistic than those considered in prior work. Our approach, \emph{approximately co-sufficient sampling via Bayes} , considerably expands the scope of this flexible type of goodness-of-fit testing. We prove the approximate validity of the resulting test, and demonstrate its utility on three common null models where no existing methods apply, as well as its outperformance on models where existing methods do apply.
\end{abstract}

\section{Introduction}

Goodness-of-fit (GoF) testing refers to the problem of testing whether a particular family of distributions (the ``model'') is consistent with the observed data: for instance, does the data follow a Gaussian distribution? GoF testing is heavily studied in statistics and has applications across domains, including biology~\citep{guo1992performing}, economics~\citep{cowell2009goodness}, astronomy~\citep{acharya2024spectral}, and finance~\citep{frezza2014goodness}. In this paper we will consider \emph{parametric} null hypotheses of the form
\begin{equation}\label{eq: H_0}
    H_0: X \sim f_\theta \quad \textnormal{for some } \theta \in \Theta,
\end{equation}
where $\{f_\theta : \theta \in \Theta \subseteq \mathbb{R}^d\}$ represents a parametric family of densities. This is hypothesis testing with a \emph{composite} null hypothesis space (i.e., the model whose goodness-of-fit is being tested). In GoF testing, it is common to leave the alternative hypothesis unspecified in general: we can interpret the test as asking whether the model appears to fit the data, or not. However, in specific settings, it may be the case that we design a test with a particular alternative in mind, as we will see in the examples developed later on. A key challenge of GoF testing problems is that often, any alternative hypothesis of interest would typically be very high-dimensional or even infinite-dimensional: for instance, if the data does not follow a Gaussian distribution (the null), then perhaps it instead follows some heavier-tailed distribution (a nonparametric alternative).
In such settings, any powerful test statistic is often too complex to permit any theoretical calculation of its null distribution, even asymptotically.

If the null hypothesis were simple, i.e., if $\Theta = \{\theta_0\}$, then any function $T$ of the data $X$ could be used as a test statistic and converted to a valid p-value by fixing a positive integer $M$, generating i.i.d. samples $\widetilde{X}^{(1)}, \dots, \widetilde{X}^{(M)}$ from $f_{\theta_0}$, and computing
\begin{equation}\label{eq: pval}
\operatorname{pval}_T\left(X, \widetilde{X}^{(1)}, \ldots, \widetilde{X}^{(M)}\right) = \frac{1 + \sum_{m=1}^M \mathbbm{1}\{T(\widetilde{X}^{(m)}) \geq T(X)\}}{M + 1}.    
\end{equation}
To see why the above p-value is valid, note that the numerator of~\eqref{eq: pval} is the rank of $T(X)$ among $T(X),T(\widetilde{X}^{(1)}), \dots, T(\widetilde{X}^{(M)})$, and under $H_0$, these $M+1$ random variables are exchangeable.
Unlike standard parametric tests like the likelihood ratio, Wald, and score tests, which each prescribe a specific test statistic based on an alternative hypothesis space that must satisfy certain strong regularity conditions, the absolute flexibility in the choice of test statistic function $T$ in~\eqref{eq: pval} allows the user to leverage any domain knowledge, prior information, and believed structure under the alternative (no matter what it is) to make the test as powerful as possible.

The appealing strategy of the previous paragraph works because $H_0$ is a simple null---it contains just one distribution, and so the analyst knows what distribution to sample from in order to get exchangeable copies of the data. The idea of \emph{co-sufficient sampling} extends this strategy to \emph{composite} null hypotheses by conditioning on a sufficient statistic for the unknown parameter under the null model, rendering the data distribution conditionally parameter-free under $H_0$ so the analyst knows once again what \emph{conditional} distribution to sample from in order to get exchangeable copies of the data. 
Exact co-sufficient sampling \citep{bartlett1937properties, engen1997stochastic,agresti1992survey,stephens2012goodness} can only be fruitfully applied for a very narrow class of parametric null models, motivating approximate versions \citep{acss,reg_acss,xie2025generalized,awan2020approximate} that apply to a much broader class of models. Yet these existing works' limitations in scope, power, and computational efficiency still prevent the idea of co-sufficient sampling from realizing its full methodological potential in terms of generality and performance.
Section~\ref{sec: preliminaries} will review in more detail the landscape of existing methods based on the idea of co-sufficient sampling, as well as their shortcomings and other related work.

\emph{Our contributions.}\label{sec: our contribution}
This work presents a novel approach to approximate co-sufficient sampling that uses draws from a Bayesian posterior distribution as the approximate sufficient statistic that is conditioned on; we refer to our method as \emph{approximate co-sufficient sampling via Bayes} (aCSS-B). We prove approximate exchangeability for aCSS-B's samples, and as a corollary, prove approximate type-I error control when the p-value \eqref{eq: pval} is constructed with its samples. Note that while our method uses Bayesian sampling as a tool, these results provide frequentist guarantees that do not rely on an assumed prior. We demonstrate its performance on a suite of examples. The main advantages of aCSS-B over prior work are that it applies considerably more broadly than previous methods and, even when previous methods apply, aCSS-B is often more powerful, and may be more straightforward to implement and more computationally efficient. We demonstrate aCSS-B's improved generality for three canonical parametric null models in which no prior methods apply: a group-sparse linear model~(\ref{sec: simulation group sparsity}), a low-rank matrix model~(\ref{sec: simulation rank one}), and a linear spline model~(\ref{sec: simulation linear spline}) (and to complement these findings, additional experiments show that aCSS-B performs well relative to existing methods on examples where prior methods do apply).

\section{Background} \label{sec: preliminaries}
\subsection{Overview: inference via approximate exchangeability}
Our objective is to construct a valid and powerful test of the composite parametric null hypothesis \eqref{eq: H_0}, and we reduce this problem to one of sampling (approximately) exchangeable copies $\widetilde{X}^{(1)},\dots,\widetilde{X}^{(M)}$ of the data $X$ under $H_0$, as once this is accomplished~\eqref{eq: pval} provides a valid p-value (for any test statistic function $T$). Note that for such a p-value to also provide \emph{powerful} inference, we need sufficient ``diversity" among the sampled copies; for instance, if we set $\widetilde{X}^{(m)}=X$ for all $m$, then trivially the copies would be exchangeable, but the p-value \eqref{eq: pval} would be deterministically equal to 1 for any choice of test statistic (i.e., valid but powerless). Next, we review some background on methods that use this type of strategy for inference.

\subsection{Co-sufficient sampling}\label{subsec: css}
Co-sufficient sampling (CSS) \citep{bartlett1937properties, engen1997stochastic,agresti1992survey,stephens2012goodness}  identifies a sufficient statistic $S(X)$ for the null model $H_0$ and then samples the copies $\widetilde{X}^{(m)}$ i.i.d. (and independent of $X$) from the conditional distribution of $X\mid S(X)$, denoted $f_\theta(\cdot\mid S(X))$, which by definition of sufficiency is a known conditional distribution (does not depend on $\theta$) under $H_0$. It is then immediate that $X,\widetilde{X}^{(1)},...,\widetilde{X}^{(M)}$ are conditionally i.i.d., and hence exchangeable, under $H_0$.

However, for this method to be powerful, $S(X)$ should not contain too much information about $X$, otherwise conditioning on it will force all the sampled copies to be so similar to $X$ that the p-value \eqref{eq: pval} has little or no power under the alternative (with the extreme worst case being $S(X)=X$). Unfortunately, in many cases there does not exist any such sufficient statistic; see \cite{acss} for examples where this issue arises, which include logistic regression, curved exponential families (even as simple as two independent Gaussians with equal means and unequal variances), heavy-tailed distributions, and latent variable models.

\subsection{Approximate co-sufficient sampling}\label{subsec: aCSS}

\emph{Approximate co-sufficient sampling} (aCSS) was introduced in \cite{acss} to address the limitations of CSS testing. The key idea is to identify a statistic $S(X)$ that is \emph{nearly} sufficient, in the sense that $f_\theta(\cdot\mid S(X))$ depends very weakly on $\theta$, while still not revealing too much information about $X$. Then using copies $\widetilde{X}^{(m)}$ sampled from $f_{\hat{\theta}}(\cdot\mid S(X))$, for some estimator $\hat{\theta}$, results in approximate validity due to approximate sufficiency of $S(X)$ and high power due to the the randomness left in the distribution of $X$ conditioned on $S(X)$. In particular, \cite{acss} propose using, for both $S(X)$ and $\hat{\theta}$, the maximizer of the null log likelihood plus a random linear perturbation. They prove approximate validity under conditions resembling those for asymptotic normality of the maximum likelihood estimator (MLE), and empirically demonstrate aCSS's high power on a range of examples.

Several extensions of the aCSS method have been proposed to handle more complex settings---in particular, settings where due to high dimensionality or non-regularity of the null model, the MLE is inconsistent.
These extensions enable adding regularization, via either constraints or a penalty on $\theta$, in addition to the random perturbation to the log likelihood in defining $S(X)$ and $\hat{\theta}$ in aCSS. Concretely, \cite{reg_acss} develop a version of aCSS that allows for linear constraints $a_i^\top\theta\leq b_i$ (or regularization via a corresponding penalty function, $\max_i a_i^\top\theta$, which includes settings such as the Lasso). \cite{xie2025generalized} extend further to allow for a group-wise penalty of the form $\sum_j \rho(\|\theta_{G_j}\|_2)$, where $\rho$ is smooth (e.g., the group Lasso); their work also allows regularization via nonlinear constraints, $G_i(\theta)\leq b_i$, for functions $G_i$ that are either smooth or are an $\ell_p$ norm.

\emph{Limitations of existing aCSS methods.}
The existing methods in the aCSS family---the original method proposed by \citet{acss}, and the extensions to regularized versions of aCSS developed by \citet{reg_acss} and \citet{xie2025generalized}---all suffer from certain limitations in scope. A key limitation is that these methods all require the null model to be open and convex (i.e., $\Theta\subseteq\mathbb{R}^d$ is open and convex), excluding models with any kind of structural constraint such as sparsity or low rank. While structures such as sparsity can sometimes be encoded via regularization, this is a special case and is not true for many other types of structure: for instance, none of the existing variants can encode a low-rank matrix constraint (we will consider such an example further, below). Finally, in practice even when one of these methods can be applied to a given problem, it can be computationally expensive and sensitive to tuning parameters, and can have low power.

\subsection{Additional related work}

There is an enormous literature on GoF testing dating back many decades, with classical examples including the $\chi^2$, score, likelihood ratio, and Wald tests \citep[see, e.g., GoF textbooks such as][]{d2017goodness}. GoF testing continues to be a subject of contemporary research, with recent papers considering challenging parametric or nonparametric null hypotheses and innovative tests \citep[see, e.g.,][]{candes2018panning, berrett2020conditional, lundborg2022conditional,ramdas2022testing,gangrade2023sequential,sen2014testing,saha2024robust,chwialkowski2016kernel}.
What sets CSS and aCSS type methods, including the method proposed in this paper, apart from the rest of the GoF testing literature is their flexibility in the choice of test statistic: CSS and aCSS methods are \emph{wrapper} methods in the sense that they can wrap around \emph{any} test statistic, in principle enabling them to achieve high power for many different alternative distributions via unrestricted alternative-specific choices of test statistic.

A popular way to perform resampling based tests is a parametric bootstrap style procedure, in which resamples are generated from $f_{\widehat\theta}$, where $\widehat\theta$ is an estimator of $\theta_0$.
Although such a procedure generally does not enjoy finite-sample validity guarantees and is known to fail in some settings (see, e.g., \cite{acss}), it often performs well empirically and is widely used in practice in many fields such as in propensity-score analyses~\citep{adusumilli2018bootstrap} and population genetics~\citep{emerson2001selection}.
One work closely related to the aCSS literature is \cite{awan2020approximate}, which proposes a sampling method that approximates co-sufficient sampling. However, they do not prove their sampling can be used for approximately valid testing, and it is unclear how to use their approximation error bounds to do so. 

A final point is that the method we propose in this paper relies heavily on ideas from Bayesian sampling such as Markov chain Monte Carlo (MCMC)~\citep{metropolis_hastings,gibbs_sampling} and the Laplace approximation~\citep{shun1995Laplace}, though we emphasize that our problem statement, method, and guarantees remain purely frequentist in nature.

\section{Main results} \label{sec: acss-b method}

\subsection{Method}

At a high level, any p-value of the form~\eqref{eq: pval} relies on constructing copies $\widetilde{X}^{(1)},\dots,\widetilde{X}^{(M)}$ that are, at least approximately, exchangeable with the data $X$ under the null hypothesis of goodness-of-fit. Like the original aCSS method and many related approaches, our goal in aCSS-B is to provide a mechanism for sampling these copies.

To quantify the notion of approximate exchangeability we introduce the following definition. 
\begin{Definition}[Distance to exchangeability]
    For any integer $k \geq 1$ and any set of random variables $A_1, \ldots, A_k$ with a joint distribution, define
$$
\textnormal{d}_{\textnormal{exch }}\left(A_1, \ldots, A_k\right)=\inf \left\{\textnormal{d}_{\textnormal{TV}}\left(\left(A_1, \ldots, A_k\right),\left(B_1, \ldots, B_k\right)\right): B_1, \ldots, B_k \text { are exchangeable }\right\} .
$$
Here $\textnormal{d}_{\textnormal{TV}}$ denotes the total variation distance, and the infimum is taken over all sets of $k$ random variables $B_1, \ldots, B_k$ with an exchangeable joint distribution.
\end{Definition} 
This definition allows us to guarantee approximate validity of the p-value. Indeed,
by construction, we have
\begin{equation}\label{eqn:how_to_use_dexch}\mathbb{P}\left(\operatorname{pval}_T\left(X, \widetilde{X}^{(1)}, \ldots, \widetilde{X}^{(M)}\right)\leq \alpha\right) \leq \alpha + \textnormal{d}_{\textnormal{exch }}\left(X, \widetilde{X}^{(1)}, \ldots, \widetilde{X}^{(M)}\right),\end{equation}
for any $\alpha\in[0,1]$. This is because, for any exchangeable random variables $(B_0,B_1,\dots,B_M)$, it holds that $\mathbb{P}(\operatorname{pval}_T(B_0,B_1,\dots,B_m)\leq \alpha)\leq \alpha$, and therefore,
\[\mathbb{P}\left(\operatorname{pval}_T\left(X, \widetilde{X}^{(1)}, \ldots, \widetilde{X}^{(M)}\right)\leq \alpha\right) \leq \alpha+\textnormal{d}_{\textnormal{TV}}\left(\left(X, \widetilde{X}^{(1)}, \ldots, \widetilde{X}^{(M)}\right),(B_0,B_1,\dots,B_M)\right).\]


To construct copies $\widetilde{X}^{(1)}, \ldots, \widetilde{X}^{(M)}$ that are approximately exchangeable with $X$, under the null hypothesis of goodness-of-fit, the aCSS-B procedure operates as follows: after sampling the data $X$ (assumed to be drawn from the density $f_{\theta_0}$, for some unknown $\theta_0\in\Theta$), we first define a prior density $\pi$ on $\Theta$ and generate $B$ draws from the corresponding posterior, denoted by $\widehat\theta_1,\dots,\widehat\theta_B$. We then estimate the distribution of $X\mid(\widehat\theta_1,\dots,\widehat\theta_B)$ (note that we cannot compute this conditional distribution exactly, since $\theta_0$ is unknown), and sample the copies $\widetilde{X}^{(1)},\dots,\widetilde{X}^{(M)}$ from this estimated distribution.

Now we turn to calculating the required components of the procedure. Formally, we will assume that each density $f_\theta$ in our parametric family is a density with respect to some common base measure $\nu_{\mathcal{X}}$ on $\mathcal{X}$, and will choose a prior with density $\pi$ with respect to a base measure $\nu_\Theta$ on $\Theta$. The posterior distribution of $\theta\mid X$ is then defined by the density
\begin{equation}\label{eqn:posterior_density}\pi(\theta\mid x) = \frac{\pi(\theta) f_\theta(x)}{\bar{f}_\pi(x)},\end{equation}
which is again a density with respect to $\nu_\Theta$, where
\[\bar{f}_\pi(x) = \int_\Theta \pi(\theta)f_\theta(x)\;\mathsf{d}\nu_\Theta(\theta)\]
denotes the marginal density of $X$, under the prior $\theta\sim \pi$.
(Formally, to ensure that future quantities will be well-defined, we will assume from this point on that the support of $f_\theta(x)$ is contained in the support of $\bar{f}_\pi(x)$, for every $\theta$---that is, $\bar{f}_\pi(x)$ is positive for any value $x$ we might observe. For instance, if $f_\theta$ has the same support for every $\theta$, then this assumption is satisfied.)

Next, a straightforward calculation shows that the conditional distribution of $X\mid(\widehat\theta_1,\dots,\widehat\theta_B)$ has density
\begin{equation}\label{eqn:conditional_density_X_exact}g_\pi^{\theta_0}(x\mid \widehat\theta_{1:B})\propto f_{\theta_0}(x) \cdot \prod_{b=1}^B \frac{f_{\widehat\theta_b}(x)}{\bar{f}_\pi(x)}\end{equation}
with respect to $\nu_{\mathcal{X}}$. Since $\theta_0$ is unknown, however, we will replace $f_{\theta_0}(x)$ with $\bar{f}_\pi(x)$, so that the copies $\widetilde{X}^{(m)}$ are sampled according to  density
\begin{equation}\label{eqn:conditional_density_X_approx}g_\pi(x\mid \widehat\theta_{1:B}) \propto  \frac{\prod_{b=1}^B f_{\widehat\theta_b}(x)}{\bar{f}_\pi(x)^{B-1}}.\end{equation}
These steps describe the process of generating the copies $\widetilde{X}^{(1)},\dots,\widetilde{X}^{(M)}$ for the aCSS-B procedure; see Algorithm \ref{alg: acss-b_iid_sampling} for a complete definition of the method.

\begin{algorithm}[ht!]
\caption{aCSS-B method}
\label{alg: acss-b_iid_sampling}

\KwGiven{prior density $\pi$ on $\Theta$, and test statistic $T:\mathcal{X}\to\mathbb{R}$.}
\KwData{Observe data $X \sim f_{\theta_0}$.}
Generate $B$ posterior samples,
    \[\widehat{\theta}_1, \ldots, \widehat{\theta}_B \,\mid\, X \  \overset{\textnormal{i.i.d.}}{\sim} \ \pi(\,\cdot \mid X).\]

Generate $M$ copies of the data,
    \[\widetilde{X}^{(1)}, \dots, \widetilde{X}^{(M)} \,  \mid \, X, \widehat{\theta}_{1:B} 
 \ \overset{\textnormal{i.i.d.}}{\sim} \ g_\pi(\,\cdot \mid \widehat{\theta}_{1:B}),\]
    where the density $g_\pi(\,\cdot\mid \widehat\theta_{1:B})$ is defined as
    \[g_\pi(x\mid \widehat\theta_{1:B})\propto \frac{\prod_{b=1}^B f_{\widehat{\theta}_b}(x)}{\bar{f}_\pi(x)^{B-1}}. \]

Compute the p-value 
    \[\operatorname{pval}_T\left(X, \widetilde{X}^{(1)}, \ldots, \widetilde{X}^{(M)}\right) = \frac{1 + \sum_{m=1}^M \mathbbm{1}\{T(\widetilde{X}^{(m)}) \geq T(X)\}}{M + 1}.\]
\end{algorithm}

Before presenting our finite-sample theoretical results, we first give some intuition for why the aCSS-B procedure can be expected to provide approximately exchangeable copies of the data $X$. 

\emph{Intuition: the Bayesian setting.}Suppose that we are in a true Bayesian setting, where the unknown parameter $\theta_0$ was in fact drawn from the prior density $\pi$. In that case,~\eqref{eqn:conditional_density_X_exact} gives the conditional density of $X\mid(\theta_0,\widehat\theta_1,\dots,\widehat\theta_B)$. But after marginalizing over $\theta_0$,~\eqref{eqn:conditional_density_X_approx} is the conditional density of $X\mid(\widehat\theta_1,\dots,\widehat\theta_B)$---this is the exact, rather than approximate, conditional distribution for $X$, and thus drawing the copies $\widetilde{X}^{(m)}$ from this density leads to exact exchangeability, and validity of the test.

\emph{Intuition: sufficiency of the posterior.} We will now consider an alternative viewpoint on the intuition behind the method, without assuming a Bayesian setting---that is, we return to the setting of a fixed $\theta_0$. After observing $B$ draws from the posterior (for a large $B$), we have approximately observed the posterior density of $\theta\mid X$ with respect to the base measure $\nu_\Theta$, which is computed in~\eqref{eqn:posterior_density}.
The following standard result, proved for completeness in Appendix~\ref{app: proof of prop: posterior is sufficient}, tells us that $\pi(\,\cdot \mid X)$, which we view as statistic of the data (i.e., a map from the data $X$ to this density function), is in fact a minimal sufficient statistic.

\begin{proposition}\label{prop: posterior is sufficient}
Let $X\sim f_\theta$ for the parametric family $\{f_\theta : \theta\in\Theta\}$. Fix any prior on $\Theta$, with a positive density $\pi$ (relative to some base measure $\nu_\Theta$). Then the posterior density $\pi(\,\cdot \mid X)$ is a minimal sufficient statistic for $X$.
\end{proposition}

\noindent In other words, by conditioning on $\widehat\theta_1,\dots,\widehat\theta_B$, we are conditioning on an approximation to the posterior distribution, which can be approximated arbitrarily well by the empirical measure of its samples~\citep{Vapnik2013}. Since the posterior distribution is a sufficient statistic, this means that the copies $\widetilde{X}^{(m)}$ will be approximately exchangeable with $X$ (for large $B$), since we have nearly removed the effect of the unknown parameter $\theta_0$.

\emph{Comparison to aCSS and its extensions.}
The core idea of our method is similar to the aCSS method of \cite{acss} (and the regularized extensions of aCSS proposed by \citet{reg_acss} and \citet{xie2025generalized}), since our idea
is to condition on an approximately sufficient statistic for $X$ in order to generate the copies $\widetilde{X}^{(1)},\ldots,\widetilde{X}^{(M)}$.
However, our method makes a very different choice of information to condition on: while aCSS and its regularized variants each condition on a noisy version of the MLE for $\theta_0$ given $X$ (or, a noisy version of the regularized MLE), we instead condition on a large number $B$ of draws from the posterior distribution of $\theta\mid X$. This different approximate sufficient statistic only requires us to sample from the posterior distribution of $\theta\mid X$, as opposed to requiring optimization of the likelihood, which can be computationally more expensive in many problems~\citep{ma2019sampling}. We will also see in our empirical examples below that aCSS-B offers greater flexibility and is applicable to a wider range of problems.

\subsection{Theoretical Guarantees}\label{sec: theoretical guarantees}
As we have seen in Proposition \ref{prop: posterior is sufficient}, the posterior density $\pi(\,\cdot\mid X)$ is a minimal sufficient statistic for the data $X$; this explains why, as $B\to\infty$, we can expect that the aCSS-B method should be valid. In practice, of course, we run aCSS-B with a finite set of posterior samples---leading to approximately exchangeable $(X, \widetilde{X}^{(1)}, \dots, \widetilde{X}^{(M)})$---thereby yielding an approximately valid test. 

In this section, we examine the behaviour of the method when $B$ is finite, to understand the role of $B$ in the validity of the aCSS-B approach, for which we will first need a few definitions. Below, let $\pi_0$ denote any prior density on $\Theta$, with respect to the same base measure $\nu_\Theta$. For intuition, we should think of $\pi_0$ as being concentrated near the unknown true parameter value $\theta_0$. 

\begin{Definition}[Prior concentration]\label{def:definition_1}
    For any prior with density $\pi_0$,
    define 
\begin{align}
\epsilon(\pi_0) = \textnormal{d}_{\textnormal{TV}}\left(f_{\theta_0}, \bar{f}_{\pi_0}\right),
\end{align}
where as before, $\bar{f}_{\pi_0}$ denotes the density of the marginal likelihood of $X$ when we draw $\theta\sim \pi_0$, i.e.,
\[\bar{f}_{\pi_0}(x) = \int_\Theta f_\theta(x)\pi_0(\theta)\;\mathsf{d}\nu_\Theta(\theta).\]
\end{Definition}
To give a concrete example, we can consider taking $\pi_0$ to be the restriction of $\pi$ to some small neighborhood around $\theta_0$---that is, $\pi_0$ is the distribution of $\theta\sim\pi$ conditional on the event $\|\theta-\theta_0\|_2\leq r$. If $r$ is sufficiently small, we would expect $\epsilon(\pi_0)$ to be small as well.
\begin{Definition}[Posterior sensitivity]\label{def:definition_2}
Given observed data $X\in\mathcal{X}$, let $\pi(\,\cdot\mid X)$ (respectively, $\pi_0(\,\cdot\mid X)$) denote the posterior distribution of $\theta$ under the prior $\theta \sim \pi$ (respectively, $\theta\sim \pi_0$). Define
\begin{align}
\Delta(\pi_0) =\mathbb{E}_{\theta_0}\left[\textnormal{d}_{\chi^2}\left(\pi_0(\,\cdot\mid X) \,\middle\|\, \pi(\,\cdot\mid X)\right)^{1 / 2}\right],
\end{align}
where $\textnormal{d}_{\chi^2}$ denotes the $\chi^2$ divergence between distributions, and where the expected value is taken with respect to $X\sim f_{\theta_0}$.
\end{Definition}
\noindent Note that, if the data $X$ carries strong information for inferring the parameter $\theta$, we might expect that the posterior is not affected too much by the choice of prior---and therefore $\Delta(\pi_0)$ would not be too large, even for some $\pi_0$ that is strongly concentrated near $\theta_0$ (so that $\epsilon(\pi_0)\approx 0$). 

We emphasize that this prior $\pi_0$ will appear in the upper bound on the type-I error in Theorem~\ref{thm: validity}, but does \emph{not} need to be specified for running the algorithm.

With the above definitions, we state the main result, which bounds the distance to exchangeability---and therefore, the type-I error of the aCSS-B procedure.

\begin{Theorem} \label{thm: validity}
    After observing the data $X$, let $\widetilde{X}^{(1)}, \ldots, \widetilde{X}^{(M)}$ be sampled as in Algorithm \ref{alg: acss-b_iid_sampling}, for some positive prior density $\pi$ on $\Theta$. Then, if $X \sim f_{\theta_0}$ for some $\theta_0 \in \Theta$,

$$
\textnormal{d}_{\textnormal{exch}}\left(X, \widetilde{X}^{(1)}, \ldots, \widetilde{X}^{(M)}\right) \leq \inf_{\pi_0}\left\{\epsilon(\pi_0)+ \frac{\Delta(\pi_0)}{2\sqrt{B}}
\right\},$$
where the infimum is taken over all densities $\pi_0$ on $\Theta$ with respect to base measure $\nu_\Theta$ such that the support of $\bar{f}_{\pi_0}(x)$ contains the support of $f_\theta(x)$ for all $\theta$. 
\end{Theorem} 
The proof of this theorem is given in appendix~\ref{app: proof of thm: validity}.
\Cref{thm: validity} states that the copies $\widetilde{X}^{(1)}, \ldots, \widetilde{X}^{(M)}$ are approximately exchangeable with $X$.
In particular, applying~\eqref{eqn:how_to_use_dexch}, this implies that for any predefined test statistic $T: \mathcal{X} \rightarrow \mathbb{R}$ and rejection threshold $\alpha \in[0,1]$, the p-value satisfies

$$
\mathbb{P}\left(\operatorname{pval}_T\left(X, \widetilde{X}^{(1)}, \ldots, \widetilde{X}^{(M)}\right) \leq \alpha\right) \leq \alpha + \inf_{\pi_0}\left\{\epsilon(\pi_0)+ \frac{\Delta(\pi_0)}{2\sqrt{B}}\right\}.
$$

\subsection{Challenges: sampling from the posterior, and sampling the copies} \label{sec: sampling the copies}
As described in Algorithm~\ref{alg: acss-b_iid_sampling}, the application of aCSS-B to any hypothesis testing problem requires two sampling steps---first, drawing $B$ independent samples $\widehat\theta_b$ from the posterior distribution, and second, sampling $M$ independent copies $\widetilde{X}^{(m)}$ from $g_\pi(\,\cdot \mid \widehat{\theta}_{1:B})$. Both of these steps may be computationally challenging in practice, and in this section we briefly describe some solutions.

First we consider sampling the posterior draws $\widehat\theta_b$. In practice, it is often only possible to sample from the posterior with MCMC techniques, and sampling exactly i.i.d.\ draws is computationally infeasible; this is a standard challenge in Bayesian statistics. In our implementation, we employ MCMC techniques such as Gibbs sampling or Metropolis--Hastings (depending on the specific setting), along with adequate thinning to reduce autocorrelation between samples~\citep{riabiz2022optimal} and burn-in~\citep{stewart2009determining} so that the samples are more representative of the target distribution. These techniques, and the extent to which they approximate i.i.d.\ sampling from the posterior sampling, are well-studied in the Bayesian literature~\citep{gelman1995bayesian, gagniuc2017markov,barber2012bayesian}, so we do not explore their theoretical properties further here.

Our second challenge is the problem of sampling the copies $\widetilde{X}^{(1)}, \dots, \widetilde{X}^{(M)}$ i.i.d.\ from $g_\pi(\,\cdot \mid \widehat{\theta}_{1:B})$.
This is again often infeasible to perform exactly---and unlike the problem of sampling from the posterior, this is no longer a standard challenge in Bayesian statistics, since it is not typical to condition on more than one draw from the posterior. Therefore, this requires careful treatment, to ensure that any approximations we take do not invalidate our finite-sample type-I error guarantees. In appendix~\ref{app:sampling_copies}, we provide a theoretical analysis of this challenge. Specifically, we treat two key issues that arise: first, that MCMC sampling techniques introduce dependence among the copies, and second, that the marginal density $\bar{f}_\pi(x)=\int_{\Theta} \pi(\theta) f\left(x ; \theta\right) \textnormal{d} \theta$ (which appears in the denominator of $g_\pi(\,\cdot \mid \widehat{\theta}_{1:B})$) may be hard to calculate exactly in some problem settings. Our results in appendix~\ref{app:sampling_copies} establish type-I error control even when we use approximate strategies for sampling the copies, and bound the increase in type-I error when we use an approximation of $\bar{f}_\pi(x)$.

\section{Experiments} \label{sec: simulation}

\subsection{Simulation Design}

In this section, we examine the performance of aCSS-B in five simulated examples.\footnote{Code for reproducing all experiments is available at \url{https://github.com/Ritwik-Bhaduri/aCSS-B/}.} The first two examples are in settings where aCSS or its regularized extensions can be applied, and the remaining three examples represent setups that are outside the scope of these methods.

Across all five examples, we implement aCSS-B with $B=25$ posterior draws, and with $M=300$ copies $\widetilde{X}^{(m)}$ sampled for running the test. 
Informally, we find aCSS-B to be easy to tune---we (successfully) use the same default value of $B$ for all five examples, and observe that standard, default priors work well out-of-the-box in each setting. 
For each example in the following subsections, details for the process of sampling the posterior draws $\widehat\theta_1,\dots,\widehat\theta_B$, and for sampling the copies $\widetilde{X}^{(1)},\dots,\widetilde{X}^{(M)}$, can be found in the corresponding subsection of appendix~\ref{app: sampling details}.

For each experiment, after computing a p-value, we test the null hypothesis at the level $\alpha = 0.05$. We compute the type-I error rate for the setting where the null hypothesis is true (that is, $X\sim f_{\theta_0}$ for some $\theta_0\in\Theta$), and compute power for settings where the null is false, across a range of different signal strengths. In each example we compare this power against an oracle, which is given access to $\theta_0$---that is, the oracle can calculate the null distribution of the test statistic $T(X)$ by simply sampling from $f_{\theta_0}$. Results are reported after averaging over $500$ independent trials, with standard error bars shown in the figures.

\subsection{Logistic Regression}\label{sec: simulation logistic regression}

\emph{The model.}
We consider the logistic regression model as in \cite{acss}. Here $X_i\in\{0,1\}$ is binary and the covariate $Z_i \in \mathbb{R}^d$ is treated as fixed; the likelihood function is
\begin{equation}\label{eqn:logistic_likelihood}
f(x ; \theta)=\prod_{i=1}^n\left(\frac{e^{Z_i^{\top} \theta}}{1+e^{Z_i^{\top} \theta}}\right)^{x_i} \cdot\left(\frac{1}{1+e^{Z_i^{\top} \theta}}\right)^{1-x_i}
\end{equation}
with parameter $\theta \in \Theta=\mathbb{R}^d$, with dimension $d=5$ and number of observations $n=100$. (We can interpret $f(x ; \theta)$ as a density with respect to the base measure $\nu_{\mathcal{X}}$ on $\mathcal{X}=\{0,1\}^n$ that places mass 1 on each point $x \in\mathcal{X}$, i.e., the counting measure.) The true parameter vector is given by $\theta_0 = 0.2\cdot\vec{1}_d$. We test a conditional independence hypothesis by considering a variable $Y_i\in \mathbb{R}$ whose conditional distribution given $Z_i$ is independent of $X_i$ under the null hypothesis, but is dependent under the alternative:
\[Y\mid (X,Z) \sim a\left(b(Z)+\beta_0^\top Z \cdot \mathbbm{1}_{X=0} + \beta_1^\top Z \cdot \mathbbm{1}_{X=1}\right),\]
where 
$a(t)= t+0.5 t^3$, $b(z)= 0.5 \sum_{j=1}^5\left(z_j\right)_{+}$, $ \beta_0=c \cdot \vec{e}_1$, and $\beta_1=c \cdot \vec{e}_5$, where $\vec{e}_j$ is the $j$th basis vector and where $c \in\{0,0.1,0.2, \ldots, 1\}$ indicates the signal strength (with $c=0$ corresponding to the null hypothesis). As in \citet[Example 1, Section 4.5.2]{acss}, we use a test statistic based on sliced inverse regression; see that paper for more details.

\emph{Can we apply existing aCSS methods?} \cite{acss} apply aCSS to this problem, and we will compare aCSS-B against the exact same implementation of aCSS as used for this problem in that paper.

\emph{Choice of prior for implementing aCSS-B.}
We choose the prior density $\pi$ as
\[\pi(\theta) = \prod_{j=1}^d \phi(\theta_j;0,1),\]
where $\phi\left(\,\cdot ; \mu_j, \sigma^2\right)$ is the density of the normal distribution with mean $\mu$ and variance $\sigma^2$---that is, $\pi$ is a standard Gaussian prior on the parameter vector 
$\theta\in\mathbb{R}^d$.

\emph{Results.}
\begin{figure}[t]
    \centering
    \includegraphics[width=0.7\linewidth]{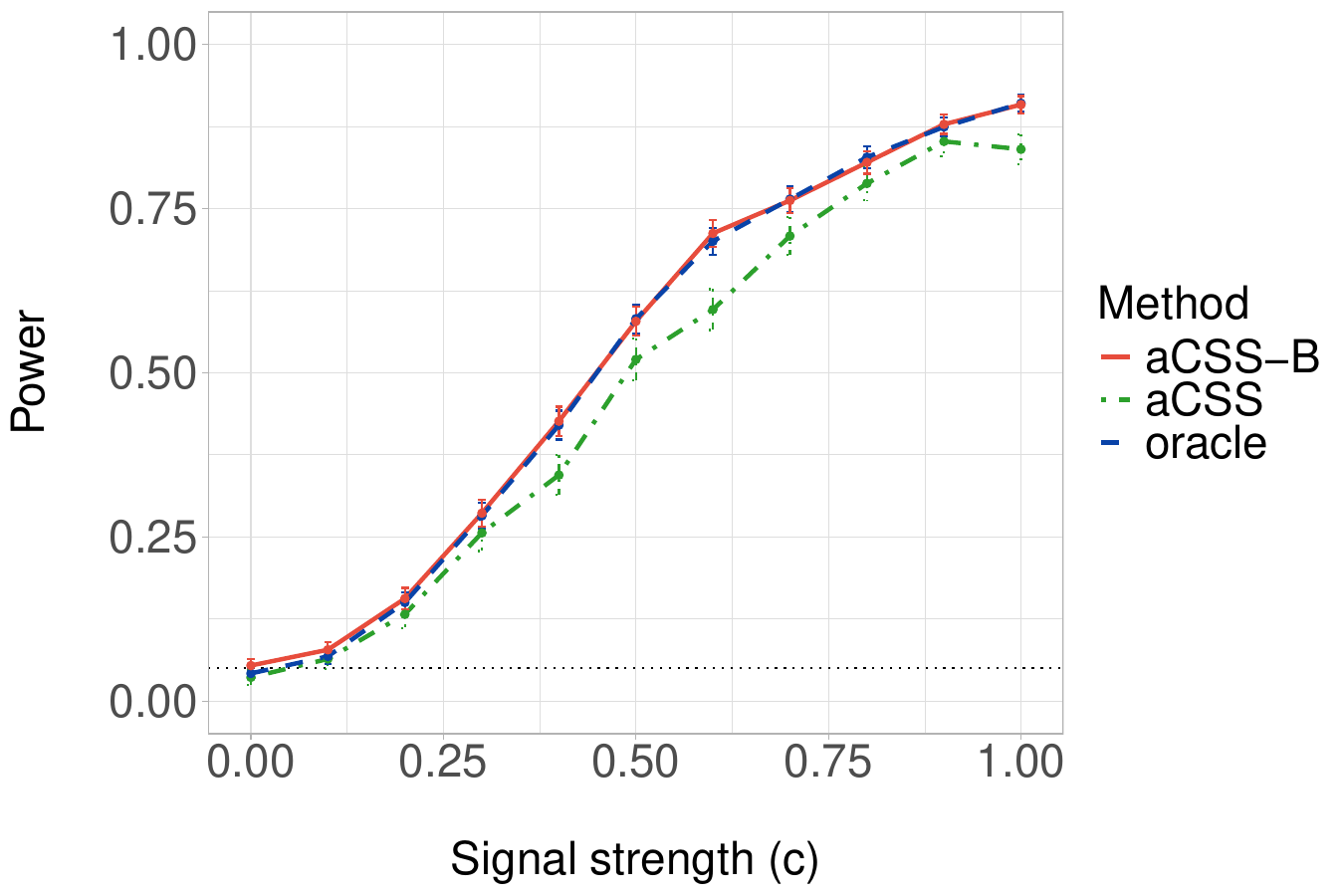}
    \caption{Power comparison between aCSS-B, aCSS, and an oracle for the logistic regression model of Section~\ref{sec: simulation logistic regression}.
    }
    \label{fig:logistic_power}
\end{figure}
Figure~\ref{fig:logistic_power} compares the performance of aCSS-B to that of aCSS and the oracle.
For this example, the oracle consists of sampling the copies $\widetilde{X}^{(m)}$ from the logistic model specified in~\eqref{eqn:logistic_likelihood}, independently of $Y$.
First, we see that all three methods result in a type-I error level of $5\%$ under the null (i.e., signal strength $c=0$). Under the alternative (signal strength $c>0$), the methods show similar power, but we can see that aCSS has slightly lower power than the oracle, while aCSS-B appears to have power equal to that of the oracle.

\subsection{Mixture of Gaussians}\label{sec: simulation Gaussian mixture}

\emph{The model.}
This example is taken from \cite{reg_acss}. The null data distribution is given by sampling $n$ i.i.d.\ draws from a mixture of two Gaussians, so that the likelihood function for the data $X = (X_1, \dots, X_n)$ is given by
$$
f(x ; \theta)=\prod_{i=1}^n \left(w_1 \phi(x_i; \mu_1,\sigma^2_1) + (1-w_1)\phi(x_i;\mu_2,\sigma^2_2)\right).
$$ This family of distributions is therefore parametrized by 
$\theta = (w_1, \mu_1, \sigma^2_1, \mu_2, \sigma^2_2) \in \Theta$, where
\[
\Theta = (0,1)\times \mathbb{R}\times \mathbb{R}_+\times \mathbb{R}\times \mathbb{R}_+ \subseteq\mathbb{R}^5.
\]
Therefore, the GoF test can be interpreted as testing the null hypothesis that the underlying model is a Gaussian mixture with $2$ components. We will consider an alternative that the data is instead drawn from a Gaussian mixture with $>2$ components. In our simulations, we take $n=200$ and
the data is generated from the following mixture
\begin{equation}\label{eqn:mixture_three_part}
p\, \mathcal{N}(0,0.01)+\frac{1-p}{2} \mathcal{N}(0.4,0.01)+\frac{1-p}{2} \mathcal{N}(-0.4,0.01)
\end{equation}
where $p = 0$ corresponds to the null hypothesis being true, while $p>0$ corresponds to the alternative. As in \citet[Section 6.1.1]{reg_acss}, we use a test statistic based on $k$-means clustering with $k=2$ versus $k=3$; see that paper for more details.

\emph{Can we apply existing aCSS methods?} For this problem, aCSS cannot be applied because the likelihood maximization problem is degenerate (since $\sigma^2_1$ or $\sigma^2_2$ can be arbitrarily close to zero, leading to a likelihood that can approach infinity). Instead, \citet{reg_acss} apply a regularized extension of aCSS (which we will refer to as reg-aCSS) to this problem, placing constraints that bound $\sigma^2_1,\sigma^2_2$ away from zero, but find low power compared to the oracle. We will compare aCSS-B against the exact same implementation of aCSS-B as used for this problem in that paper.

\emph{Choice of prior for implementing aCSS-B.} 
We assume the following prior distributions: $w_1 \sim \textnormal{Beta}(2,2)$, and 
\[\sigma_j^2\sim \textnormal{Inv-Gamma}(1,0.5), \quad \mu_j\mid \sigma_j^2 \sim \mathcal{N}(0,\sigma^2_j)\]
independently for each $j=1,2$.

\emph{Results.} 
\begin{figure}[t]
    \centering
    \includegraphics[width=0.7\linewidth]{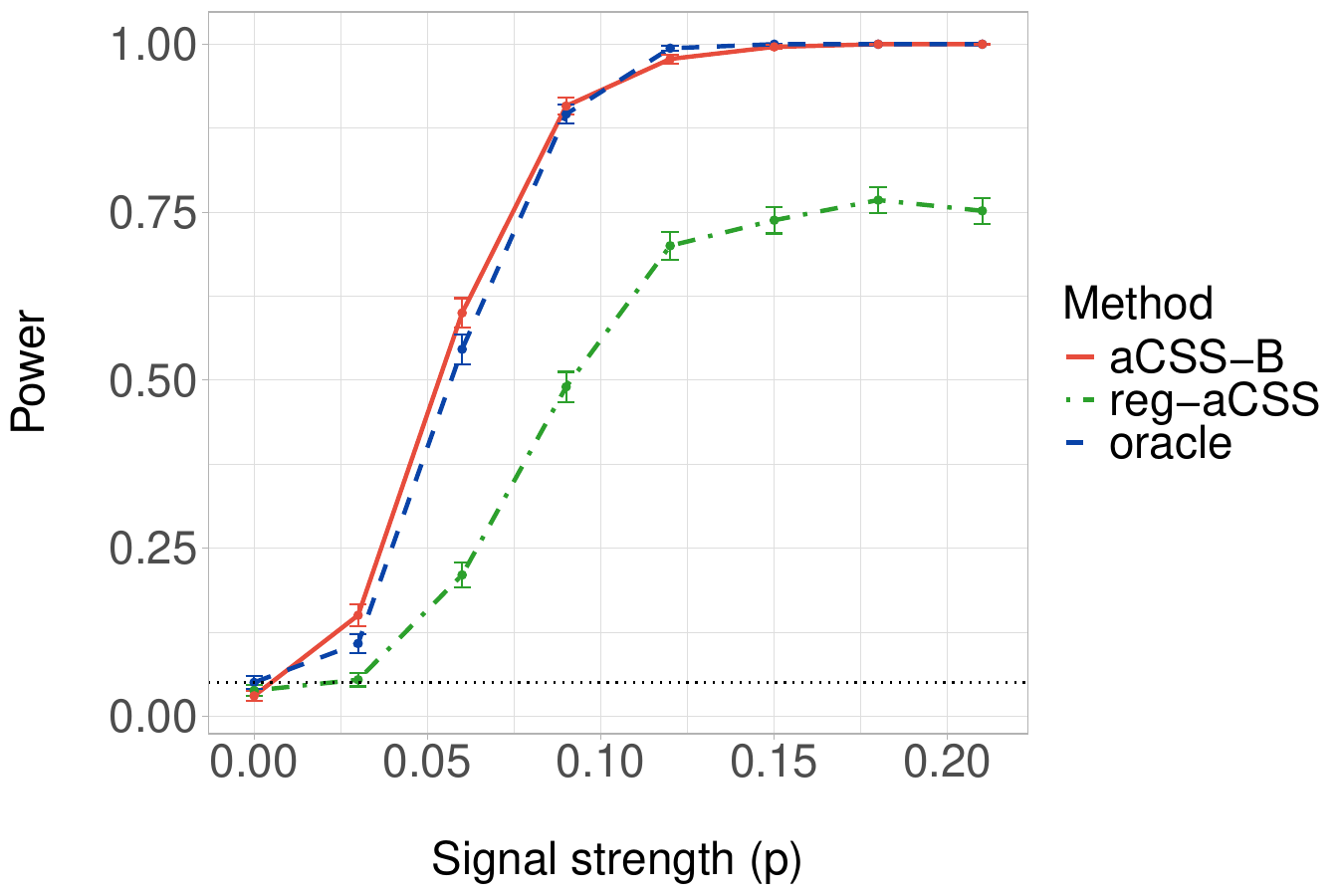}
    \caption{Power comparison between aCSS-B, reg-aCSS, and an oracle for the Gaussian mixture model of Section~\ref{sec: simulation Gaussian mixture}.
    }    
    \label{fig:Gaussian mixture power}
\end{figure}
Figure~\ref{fig:Gaussian mixture power} compares the performance of aCSS-B to that of reg-aCSS and the oracle. For this example, the oracle consists of sampling the entries of $\widetilde{X}^{(m)}$ i.i.d.\ from the two-component mixture specified in~\eqref{eqn:mixture_three_part} if we set $p=0$.
All three methods result in a type-I error level of $5\%$ under the null (i.e., mixture weight $p=0$). Under the alternative (mixture weight $p>0$), aCSS-B enjoys similar power as the oracle across nearly all values of $p$, while reg-aCSS shows lower power.

\subsection{Rank-1 matrix}\label{sec: simulation rank one}

Next we turn to examples that are beyond the scope of aCSS and its regularized extensions. Our first such example lies in the setting of low-rank matrix data.

\emph{The model.}
We assume that the data $X\in\mathbb{R}^{n\times n}$ (with $n=10$) is generated as
\[X = A + W,\]
where $A$ is a fixed matrix representing the underlying signal, while $W$ is noise, with $W_{ij}\overset{\textnormal{i.i.d.}}{\sim} \mathcal{N}(0,0.25)$. Under the null, the signal $A$ is a rank-1 matrix. Therefore the GoF test can be represented with the family
\[f_\theta(x) = \prod_{i=1}^n\prod_{j=1}^n \phi(X_{ij} - A_{ij} ; 0,0.25),\]
parametrized by $\theta = A\in\Theta\subseteq\mathbb{R}^{n\times n}$, where $\Theta$ is the space of rank-1 matrices. Under the alternative, the rank of the underlying signal is $>1$.

In order to generate the data, we define $A_0 = U_1V_1^\top + c U_2V_2^\top$ where $U = [U_1,U_2],V = [V_1,V_2]\in\mathbb{R}^{n\times 2}$ have i.i.d.\ $\mathcal{N}(0,1)$ entries. We vary c, with $c=0$ corresponding to the null---note that $\text{rank}(A)=1$ under the null and $\text{rank}(A)=2$ under the alternative. The test statistic $T(X)$ is defined as the second largest eigenvalue of $X^\top X$. 

\emph{Can we apply existing aCSS methods?} In this example, the null parameter space is $\Theta = \{A\in \mathbb{R}^{n\times n}: \textnormal{rank}(A)=1\}$. The challenging nature of the rank-1 constraint means that none of the existing aCSS methods can be applied---specifically, \citet{acss} would require a convex and open null model space, while the regularized forms of aCSS \citep{reg_acss,xie2025generalized} allow constraints on the parameter $\theta$ but only limited types of constraints are allowed, which again cannot encompass a rank-1 restriction.
(We could instead consider reparametrizing as $A=uv^\top$ and taking $\theta=(u,v)$, but the assumptions of aCSS are again violated---in this case, because the existing aCSS theory requires strong convexity of the log-likelihood at the MLE, which cannot hold under this reparametrization due to nonidentifiability.)
However, aCSS-B can be applied here with a suitable choice of prior.

\emph{Choice of prior for implementing aCSS-B.}
To specify the prior distribution for the null, we introduce two vectors $U, V\in \mathbb{R}^n$ such that $A = UV^\top $ and assume independent multivariate standard Gaussian priors for $U$ and $V$, that is, $U,V\overset{\textnormal{i.i.d.}}{\sim} \mathcal{N}(\vec{0}_n,\mathbb{I}_n)$.

\emph{Results.}
\begin{figure}[t]
    \centering
    \includegraphics[width=0.7\linewidth]{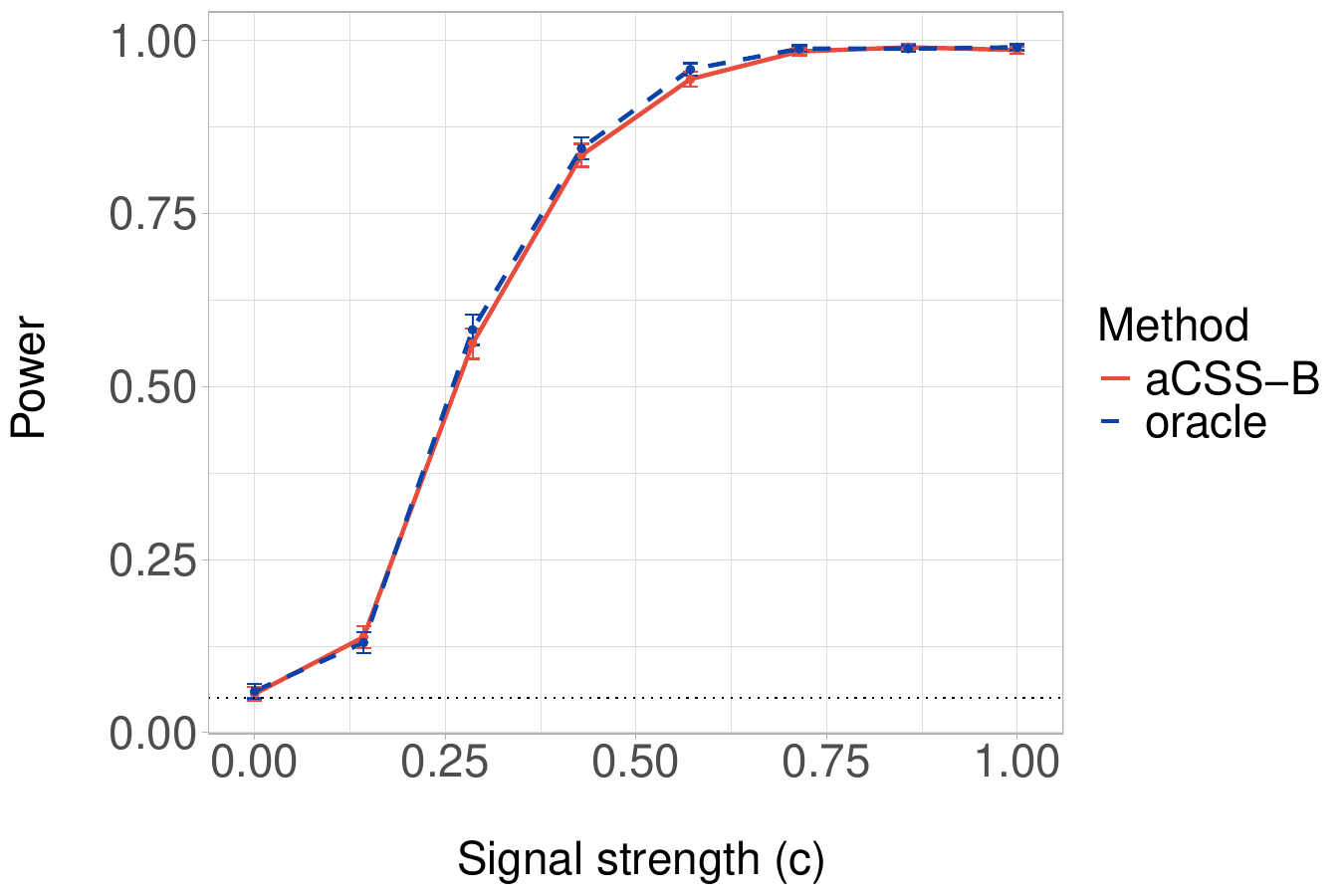}
    \caption{Power comparison between aCSS-B and an oracle for the rank-1 matrix model of Section~\ref{sec: simulation rank one}.
    }
\label{fig:rank one matrix power}
\end{figure}
Figure~\ref{fig:rank one matrix power} compares the performance of aCSS-B to that of the oracle. For this example, the oracle consists of sampling $\widetilde{X}^{(m)} = A_0+W$ where $A_0 = U_1V_1^\top$ and $W$ has i.i.d standard normal entries (recall that the data is generated with mean $A=U_1V_1^\top + cU_2V_2^\top$, where $0\leq c\leq1$, so this choice of $A_0$ represents a rank-1 approximation to the true distribution).
We can see that the aCSS-B method achieves type-I error control at level $5\%$ under the null (i.e., when $A$ has rank $1$), as desired, and has nearly the same power as the oracle under the alternative.

\subsection{Group-sparse regression}\label{sec: simulation group sparsity}

Our next example studies a group-sparse linear regression model.

\emph{The model.}
We consider the following model for data $X\in\mathbb{R}^n$, given covariates $Z\in\mathbb{R}^{n\times d}$ (which we treat as fixed):
\begin{equation}\label{eqn:group_sparse_model}
X = Z \beta + \epsilon\textnormal{ where }
\epsilon_i \overset{\textnormal{i.i.d.}}{\sim} \mathcal{N}(0, 1),
\end{equation}
where we chose $n=100$ and $d=50$. Given the partition $\{1,\dots,d\} = I_1\cup \dots \cup I_{10}$ into $G=10$ groups of equal size, $|I_g| = 5$, our null hypothesis is that only one group is ``active'' in the regression: that is, the active set $A=\{g : \beta_{I_g}\neq (0,\dots,0)\}$ has size $|A|=1$ while under the alternative $|A|>1$.

To generate the data, we draw the entries of $Z$ independently from $\mathcal{N}(0,1)$ and choose two group indices $g_1\neq g_2$ uniformly at random from the $10$ groups. The coefficients are generated as follows: 

\begin{equation}\label{eq: coeff_group_sparsity}
    \beta_{I_{g_1}} = \vec{1}_{|I_{g_1}|}, \ 
    \beta_{I_{g_2}} = c\,\vec{1}_{|I_{g_2}|}, \ \ 
    \beta_j = 0 \ \forall \ j \in [d]\backslash(I_{g_1} \cup I_{g_2}).
\end{equation}

For the null, we consider $c=0$ so that there is only one active group (namely, $g_1$), and for the alternative we take $c\in\left(0\right.,\left.0.3\right]$ so that there are two active groups ($g_1$ and $g_2$). We refer to $c$ as the signal strength.

To define the test statistic, we first compute an estimate $\widehat\beta$ of the regression coefficients
by fitting 5-fold cross-validated group LASSO from the \textnormal{R} package \texttt{gglasso}~\citep{gglasso_package} on the data. The test statistic is then defined as
$$T(X):= \frac{\sum_{g \neq \widehat g} ||\widehat\beta_{I_g}||_{\infty}}{||\widehat\beta_{I_{\widehat g}}||_{\infty}}\textnormal{ where }\widehat g = \underset{g\in [G]}\arg\max ||\widehat\beta_{I_g}||_\infty,$$
which measures the sum of the largest estimated coefficients \emph{outside} of the top selected group $\widehat{g}$ relative to the largest one within $\widehat{g}$.

\emph{Can we apply existing aCSS methods?} In this example, the null parameter space is 
$$
\Theta = \{(\beta_1, \dots, \beta_d) \in \mathbb{R}^d : \beta_{I_{g}} = \vec{0}_{d_{g}} \textnormal{ for all } g \neq g^\star, \textnormal{ for some } g^\star \in \{1,\dots, G\}\}.
$$ As for the rank-1 constraint in the previous example, aCSS and its regularized extensions cannot be applied to this problem because of the challenging nature of the group-sparsity constraint.

\emph{Choice of prior for implementing aCSS-B.}
For the prior $\pi$, we consider a discrete uniform distribution for the active group and choose a standard Gaussian prior for the coefficients of the active group, while the rest of the coefficients are set to zero:
\begin{align}
&g^\star \sim \textnormal{Unif(\{1,\dots, G\})}\nonumber,\\
&\beta_{I_{g^\star}}\sim \mathcal{N}(5\cdot \vec{1}_{|I_{g^\star}|},\mathbb{I}_{|I_{g^\star}|})\nonumber,\\
&\beta_{I_{g}} =\vec{0}_{|I_g|}\ \forall \ g \neq g^\star.\label{eq: prior_group_sparsity}
\end{align}

\emph{Results.}
\begin{figure}[t]
    \centering
    \includegraphics[width=0.7\linewidth]{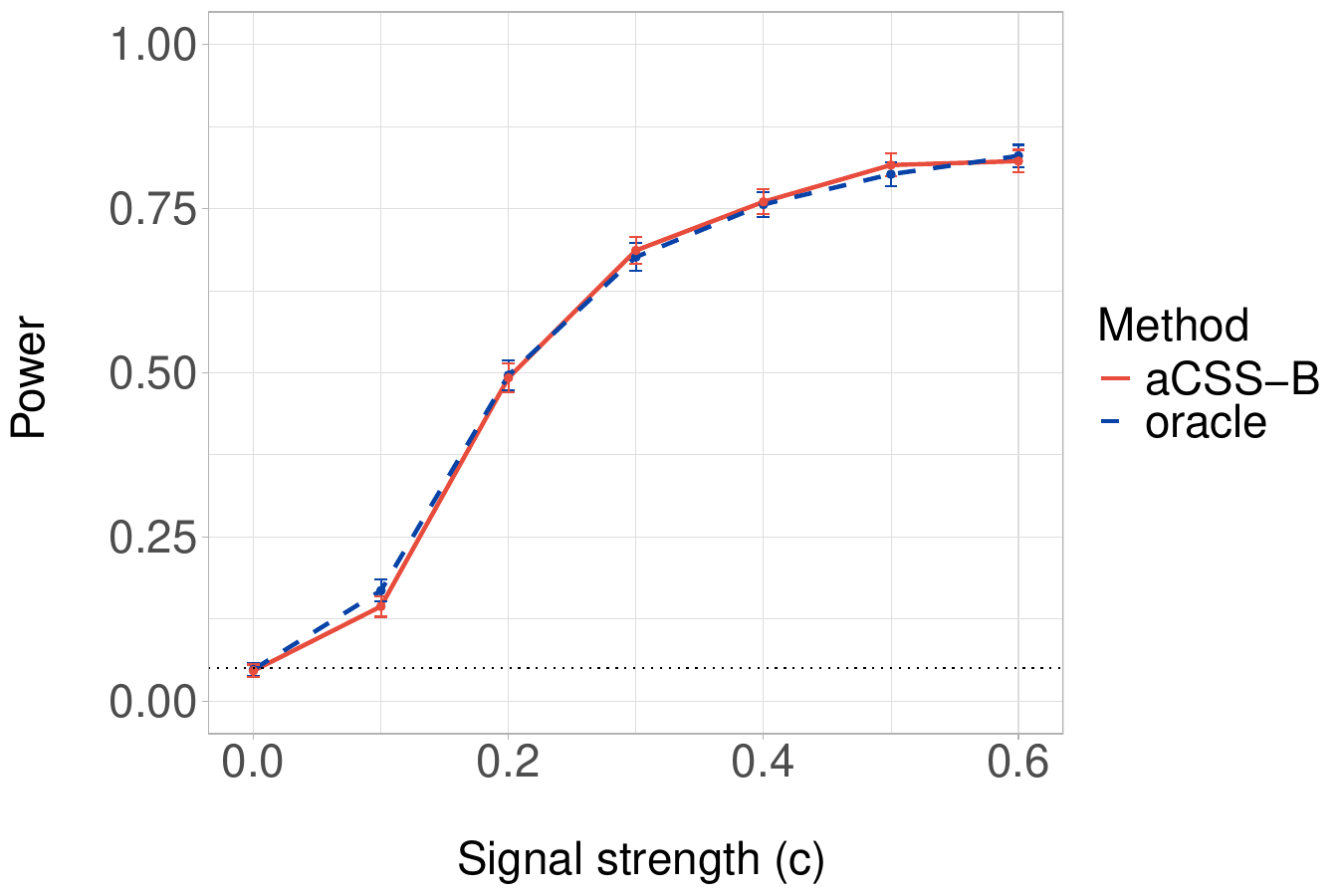}
    \caption{Power comparison between aCSS-B and an oracle for the group sparse model of Section~\ref{sec: simulation group sparsity}.}
    \label{fig:group sparsity power}
\end{figure}
Figure~\ref{fig:group sparsity power} compares the performance of aCSS-B to that of the oracle. For this example, the oracle consists of sampling $\widetilde{X}^{(m)}$ from a linear model as in~\eqref{eqn:group_sparse_model} where the coefficient vector $\beta$ defined in~\eqref{eq: coeff_group_sparsity} is redefined to have a single active group by setting $\beta_{I_{g_2}}$ to be zero (while keeping the original coefficients in the first active group, $\beta_{I_{g_1}}$).
We can see that the aCSS-B method achieves type-I error control at level $5\%$ under the null (i.e., when the coefficient vector indeed has only one nonzero group), and the power is essentially the same as the oracle under the alternative.

\subsection{Linear spline regression model}\label{sec: simulation linear spline}
Our final example is in a nonlinear regression setting, where $X$ follows a linear spline model given covariates $Z$.

\emph{The model.}
Consider a linear spline model with $k$ knots $t_1<\dots<t_k\in\mathbb{R}$, with $n=50$ observations,
\begin{align}\label{eq:linear_spline}
X_i = \mu_i + \epsilon_i \textnormal{ for } \epsilon_i \overset{\textnormal{i.i.d.}}{\sim} \mathcal{N}(0,0.25),
\end{align}
where 
\[\mu_i = \beta_{0j} + \beta_{1j}Z_i \textnormal{ if }t_{j-1}\leq Z_i< t_j.\]
(Here for convenience we
define $t_0 = -\infty$ and $t_{k+1} = \infty$; as in previous examples the covariates $Z_i\in\mathbb{R}$ are treated as fixed.) 
The parameters $\beta_{ij}\in \mathbb{R}$ are constrained so that the mean function is continuous in $\mathbb{R}$. 
The null hypothesis corresponds to the case where we have exactly $k = 1$ knot, while under the alternative we have $k=2$. 

To generate data from this distribution, we generate $Z_i \overset{\textnormal{i.i.d.}}{\sim} \text{Unif}(-5,5)$ and choose two knots as $t_1 = -1.67$ and $t_2 = 1.67$. The coefficients are generated as \begin{equation}\label{eq:linear_spline_coefs}\beta_{01}=1, \beta_{11} = -1, \beta_{21} = 1,\beta_{31} = 1-c\end{equation} and the other intercepts $\beta_{02},\beta_{03}$ are chosen so that the mean function is continuous. We consider a sequence of equally spaced values for $c$ from $0$ to $1.8$---note that $c=0$ corresponds to the null, since the slopes of the second and third segment are the same and thus effectively we only have $k=1$ knot. Under the alternative $c>0$, there are $k=2$ knots, with larger values of $c$ denoting further deviation from the null.

The test statistic $T(X)$ is the residual sum of squares from a linear spline model with one knot, which we fit using the \texttt{segmented}~\citep{segmented_package} package in \texttt{R}.

\emph{Can we apply existing aCSS methods?} In this example, the null parameter space is 
\begin{align*}
\Theta = 
&\Big\{ (\beta_{01}, \beta_{11}, \dots, \beta_{0,k+1}, \beta_{1,k+1}, t_1, \dots, t_{k}) \in \mathbb{R}^{3k+2}: \\
&\quad \beta_{0j} + \beta_{1j}t_j = \beta_{0(j+1)} + \beta_{1(j+1)}t_j \textnormal{ for all } j = 1, \dots, k , \ t_1 < \dots < t_k\Big\}.
\end{align*}
The above constraints, which stem from the continuity of the mean function at the knots in the linear spline model, cannot be accommodated by the existing aCSS methods.
However, aCSS-B can still be applied, through a carefully constructed prior as we will see below.

\emph{Choice of prior for implementing aCSS-B.}
While we can choose priors which respect the constraints in this problem, sampling from the resulting posterior distribution would be complicated. However, we can avoid this problem by using the following reparametrization:
\begin{equation} \label{eq: spline_reparametrize}
    X = \gamma_0 + \gamma_1 Z + \sum_{j=1}^{k} \gamma_{1+j} b_j(Z) + \epsilon,
\end{equation}
where  $b_j(Z) = (Z - t_j)_{+}$ is applied element-wise to $Z=(Z_1,\dots,Z_n)$, for all $j=1,\dots,k$.
In this reparametrization, $\gamma = (\gamma_0,\dots,\gamma_{k+1})\in\mathbb{R}^{k+2}$ is unconstrained, and the knots $t_1,\dots,t_k$ do not need to be ordered. Therefore, the parameter space becomes $\Theta = \mathbb{R}^{2k+2}$. (We note, however, that existing aCSS methods nonetheless cannot be applied, even with this reparametrization---this is because the log-likelihood is no longer differentiable with respect to the parameters $t_j$.)

On these unconstrained parameters, we choose the standard Gaussian priors:
\begin{align*}
&\gamma_0,\gamma_1,\gamma_2\overset{\textnormal{i.i.d.}}{\sim} \mathcal{N}(0,1),\\
&t_1\sim \mathcal{N}(0,1),
\end{align*}
for the null model with $k=1$ knots.

\emph{Results.}
\begin{figure}[t]
    \centering
    \includegraphics[width=0.7\linewidth]{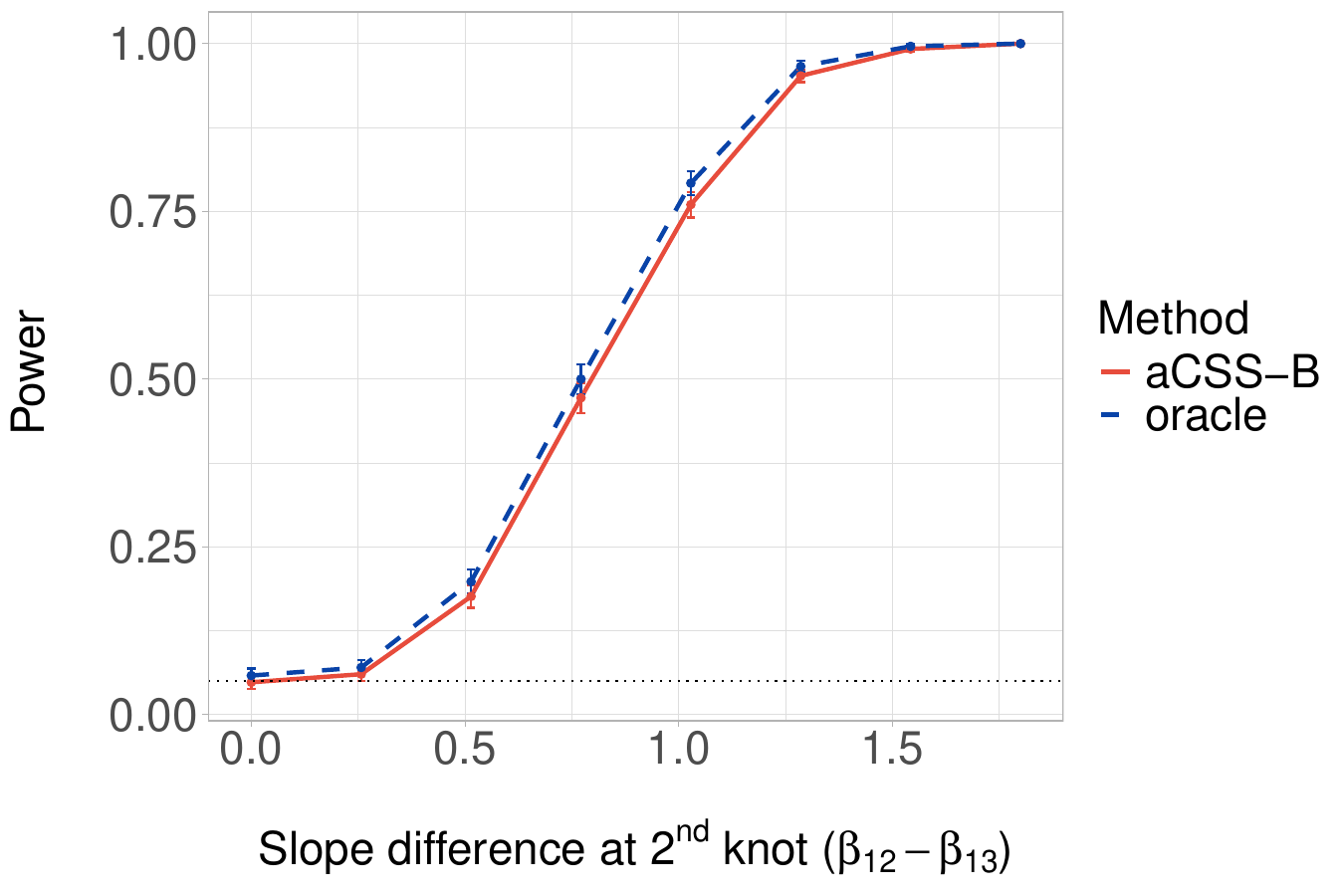}
    \caption{Power comparison between aCSS-B and an oracle for the linear spline model of Section~\ref{sec: simulation linear spline}.
}
    \label{fig: linear spline power}
\end{figure}
Figure~\ref{fig: linear spline power} illustrates the performance aCSS-B as compared to the oracle. For this example, the oracle consists of sampling $\widetilde{X}^{(m)}$'s from the model given by~\eqref{eq:linear_spline} and \eqref{eq:linear_spline_coefs} with $c=0$. We see that the aCSS-B method achieves type-I error control at level $5\%$ under the null, and the power is nearly the same as the oracle under the alternative. 

\subsection*{Acknowledgements}
R.B. and L.J. were partially supported by the National Science Foundation via grant DMS-2045981.
A.B. and R.F.B. were partially supported by the National Science Foundation via grant DMS-2023109. R.F.B. was partially supported by the Office of Naval Research via grant N00014-24-1-2544. 

\bibliographystyle{apalike}
\bibliography{reference}

@article{xie2025generalized,
  title={A Generalized Framework for Approximate Co-Sufficient Sampling},
  author={Xie, Jie and Huang, Dongming},
  journal={arXiv preprint arXiv:2506.12334},
  year={2025}
}

@book{owen_mcbook,
   author = {Art B. Owen},
   year = 2013,
   title = {Monte {C}arlo theory, methods and examples},
   publisher = {\url{https://artowen.su.domains/mc/}}
}

@book{casella2002statistical,
title={Statistical Inference},
author={Casella, George and Berger, Roger L},
year={2002},
publisher={Thomson Learning}
}

@article{fraser1963sufficiency,
  title={On sufficiency and the exponential family},
  author={Fraser, DAS},
  journal={Journal of the Royal Statistical Society: Series B (Methodological)},
  volume={25},
  number={1},
  pages={115--123},
  year={1963},
  publisher={Wiley Online Library}
}

@article{chib2001marginal,
  title={Marginal likelihood from the {M}etropolis--{H}astings output},
  author={Chib, Siddhartha and Jeliazkov, Ivan},
  journal={Journal of the American statistical association},
  volume={96},
  number={453},
  pages={270--281},
  year={2001},
  publisher={Taylor \& Francis}
}

@inproceedings{barber2016laplace,
  title={Laplace approximation in high-dimensional {B}ayesian regression},
  author={Barber, Rina Foygel and Drton, Mathias and Tan, Kean Ming},
  booktitle={Statistical Analysis for High-Dimensional Data: The Abel Symposium 2014},
  pages={15--36},
  year={2016},
  organization={Springer}
}

@article{stewart2009determining,
  title={Determining optimum burn-in and replacement times using {B}ayesian decision theory},
  author={Stewart, Leland T and Johnson, Jerry D},
  journal={IEEE Transactions on Reliability},
  volume={21},
  number={3},
  pages={170--175},
  year={2009},
  publisher={IEEE}
}

@article{riabiz2022optimal,
  title={Optimal thinning of {MCMC} output},
  author={Riabiz, Marina and Chen, Wilson Ye and Cockayne, Jon and Swietach, Pawel and Niederer, Steven A and Mackey, Lester and Oates, Chris J},
  journal={Journal of the Royal Statistical Society Series B: Statistical Methodology},
  volume={84},
  number={4},
  pages={1059--1081},
  year={2022},
  publisher={Oxford University Press}
}

@Inbook{Vapnik2013,
author={Vapnik, Vladimir N.
and Chervonenkis, Alexey Ya.},
title={On the Uniform Convergence of the Frequencies of Occurrence of Events to Their Probabilities},
bookTitle={Empirical Inference: Festschrift in Honor of Vladimir N. Vapnik},
year="2013",
publisher="Springer Berlin Heidelberg",
address="Berlin, Heidelberg",
pages="7--12",
isbn="978-3-642-41136-6",
doi="10.1007/978-3-642-41136-6_2",
url="https://doi.org/10.1007/978-3-642-41136-6_2"
}

@article{acss,
  title={Testing goodness-of-fit and conditional independence with approximate co-sufficient sampling},
  author={Barber, Rina Foygel and Janson, Lucas},
  journal={The Annals of Statistics},
  volume={50},
  number={5},
  pages={2514--2544},
  year={2022},
  publisher={Institute of Mathematical Statistics}
}

@article{reg_acss,
  title={Approximate co-sufficient sampling with regularization},
  author={Zhu, Wanrong and Barber, Rina Foygel},
  journal={arXiv preprint arXiv:2309.08063},
  year={2023}
}

@article{engen1997stochastic,
  title={Stochastic simulations conditioned on sufficient statistics},
  author={Engen, Steinar and Lilleg{\aa}rd, Magnar},
  journal={Biometrika},
  volume={84},
  number={1},
  pages={235--240},
  year={1997},
  publisher={Oxford University Press}
}

@article{shun1995laplace,
  title={Laplace approximation of high dimensional integrals},
  author={Shun, Zhenming and McCullagh, Peter},
  journal={Journal of the Royal Statistical Society Series B: Statistical Methodology},
  volume={57},
  number={4},
  pages={749--760},
  year={1995},
  publisher={Oxford University Press}
}

@Manual{segmented_package,
  title        = {segmented: Regression Models with Break-Points / Change-Points Estimation},
  author       = {Vito M. Muggeo},
  year         = {2023},
  note         = {R package version 1.6-4},
  url          = {https://CRAN.R-project.org/package=segmented}
}

@Manual{gglasso_package,
  title = {gglasso: Group Lasso Regularization},
  author = {Chunlei Wu and Kenneth Lange},
  year = {2020},
  note = {R package version 1.5},
  url = {https://cran.r-project.org/package=gglasso}
}

@article{candes2018panning,
  title={Panning for gold:‘model-X’knockoffs for high dimensional controlled variable selection},
  author={Cand{\`e}s, Emmanuel and Fan, Yingying and Janson, Lucas and Lv, Jinchi},
  journal={Journal of the Royal Statistical Society Series B: Statistical Methodology},
  volume={80},
  number={3},
  pages={551--577},
  year={2018},
  publisher={Oxford University Press}
}

@article{metropolis_hastings,
  title={Understanding the {M}etropolis--{H}astings algorithm},
  author={Chib, Siddhartha and Greenberg, Edward},
  journal={The american statistician},
  volume={49},
  number={4},
  pages={327--335},
  year={1995},
  publisher={Taylor \& Francis}
}

@article{gibbs_sampling,
  title={Explaining the {G}ibbs sampler},
  author={Casella, George and George, Edward I},
  journal={The American Statistician},
  volume={46},
  number={3},
  pages={167--174},
  year={1992},
  publisher={Taylor \& Francis}
}

@book{gelman1995bayesian,
  title={Bayesian data analysis},
  author={Gelman, Andrew and Carlin, John B and Stern, Hal S and Rubin, Donald B},
  year={1995},
  publisher={Chapman and Hall/CRC}
}

@article{agresti1992survey,
  title={A survey of exact inference for contingency tables},
  author={Agresti, Alan},
  journal={Statistical science},
  volume={7},
  number={1},
  pages={131--153},
  year={1992},
  publisher={Institute of Mathematical Statistics}
}

@article{ma2019sampling,
  title={Sampling can be faster than optimization},
  author={Ma, Yi-An and Chen, Yuansi and Jin, Chi and Flammarion, Nicolas and Jordan, Michael I},
  journal={Proceedings of the National Academy of Sciences},
  volume={116},
  number={42},
  pages={20881--20885},
  year={2019},
  publisher={National Academy of Sciences}
}

@article{acharya2024spectral,
  title={Spectral fit residuals as an indicator to increase model complexity},
  author={Acharya, Anshuman and Kashyap, Vinay L},
  journal={Research Notes of the AAS},
  volume={8},
  number={1},
  pages={1},
  year={2024},
  publisher={The American Astronomical Society}
}

@article{frezza2014goodness,
  title={Goodness of fit assessment for a fractal model of stock markets},
  author={Frezza, Massimiliano},
  journal={Chaos, Solitons \& Fractals},
  volume={66},
  pages={41--50},
  year={2014},
  publisher={Elsevier}
}

@article{cowell2009goodness,
  title={Goodness-of-fit: An economic approach},
  author={Cowell, Frank and Flachaire, Emmanuel and Bandyopadhyay, Sanghamitra},
  year={2009},
  journal={Economics Series Working Papers 444, University of Oxford, Department of Economics},
  publisher={The Toyota Centre, London School of Economics and Political Science}
}

@article{guo1992performing,
  title={Performing the exact test of {H}ardy--{W}einberg proportion for multiple alleles},
  author={Guo, Sun Wei and Thompson, Elizabeth A},
  journal={Biometrics},
  pages={361--372},
  year={1992},
  publisher={JSTOR}
}

@article{stephens2012goodness,
  title={Goodness-of-fit and sufficiency: Exact and approximate tests},
  author={Stephens, Michael A},
  journal={Methodology and Computing in Applied Probability},
  volume={14},
  pages={785--791},
  year={2012},
  publisher={Springer}
}

@article{gangrade2023sequential,
  title={A sequential test for log-concavity},
  author={Gangrade, Aditya and Rinaldo, Alessandro and Ramdas, Aaditya},
  journal={arXiv preprint arXiv:2301.03542},
  year={2023}
}

@article{lundborg2022conditional,
  title={Conditional independence testing in {H}ilbert spaces with applications to functional data analysis},
  author={Lundborg, Anton Rask and Shah, Rajen D and Peters, Jonas},
  journal={Journal of the Royal Statistical Society Series B: Statistical Methodology},
  volume={84},
  number={5},
  pages={1821--1850},
  year={2022},
  publisher={Oxford University Press}
}

@inproceedings{chwialkowski2016kernel,
  title={A kernel test of goodness of fit},
  author={Chwialkowski, Kacper and Strathmann, Heiko and Gretton, Arthur},
  booktitle={International conference on machine learning},
  pages={2606--2615},
  year={2016},
  organization={PMLR}
}

@article{awan2020approximate,
  title={Approximate Co-Sufficient Sampling for Goodness-of-fit Tests and Synthetic Data},
  author={Awan, Jordan and Cai, Zhanrui},
  journal={arXiv preprint arXiv:2006.02397},
  year={2020}
}

@article{saha2024robust,
  title={Robust likelihood ratio tests for composite nulls and alternatives},
  author={Saha, Aytijhya and Ramdas, Aaditya},
  journal={arXiv preprint arXiv:2408.14015},
  year={2024}
}

@article{bartlett1937properties,
  title={Properties of sufficiency and statistical tests},
  author={Bartlett, Maurice Stevenson},
  journal={Proceedings of the Royal Society of London. Series A-Mathematical and Physical Sciences},
  volume={160},
  number={901},
  pages={268--282},
  year={1937},
  publisher={The Royal Society London}
}

@article{besag1989generalized,
  title={Generalized {M}onte {C}arlo significance tests},
  author={Besag, Julian and Clifford, Peter},
  journal={Biometrika},
  volume={76},
  number={4},
  pages={633--642},
  year={1989},
  publisher={Oxford University Press}
}

@book{d2017goodness,
  title={Goodness-of-fit-techniques},
  author={D'Agostino, Ralph B},
  year={2017},
  publisher={Routledge}
}

@article{sen2014testing,
  title={Testing independence and goodness-of-fit in linear models},
  author={Sen, Arnab and Sen, Bodhisattva},
  journal={Biometrika},
  volume={101},
  number={4},
  pages={927--942},
  year={2014},
  publisher={Oxford University Press}
}

@article{berrett2020conditional,
  title={The conditional permutation test for independence while controlling for confounders},
  author={Berrett, Thomas B and Wang, Yi and Barber, Rina Foygel and Samworth, Richard J},
  journal={Journal of the Royal Statistical Society Series B: Statistical Methodology},
  volume={82},
  number={1},
  pages={175--197},
  year={2020},
  publisher={Oxford University Press}
}

@article{ramdas2022testing,
  title={Testing exchangeability: Fork-convexity, supermartingales and e-processes},
  author={Ramdas, Aaditya and Ruf, Johannes and Larsson, Martin and Koolen, Wouter M},
  journal={International Journal of Approximate Reasoning},
  volume={141},
  pages={83--109},
  year={2022},
  publisher={Elsevier}
}

@book{gagniuc2017markov,
  title={Markov chains: from theory to implementation and experimentation},
  author={Gagniuc, Paul A},
  year={2017},
  publisher={John Wiley \& Sons}
}

@book{barber2012bayesian,
  title={Bayesian reasoning and machine learning},
  author={Barber, David},
  year={2012},
  publisher={Cambridge University Press}
}

@article{chib1995marginal,
  title={Marginal likelihood from the {G}ibbs output},
  author={Chib, Siddhartha},
  journal={Journal of the american statistical association},
  volume={90},
  number={432},
  pages={1313--1321},
  year={1995},
  publisher={Taylor \& Francis}
}

@article{newton1994approximate,
  title={Approximate {B}ayesian inference with the weighted likelihood bootstrap},
  author={Newton, Michael A and Raftery, Adrian E},
  journal={Journal of the Royal Statistical Society Series B: Statistical Methodology},
  volume={56},
  number={1},
  pages={3--26},
  year={1994},
  publisher={Oxford University Press}
}

@article{meng1996simulating,
  title={Simulating ratios of normalizing constants via a simple identity: a theoretical exploration},
  author={Meng, Xiao-Li and Wong, Wing Hung},
  journal={Statistica Sinica},
  pages={831--860},
  year={1996},
  publisher={JSTOR}
}

@article{gelman1998simulating,
  title={Simulating normalizing constants: From importance sampling to bridge sampling to path sampling},
  author={Gelman, Andrew and Meng, Xiao-Li},
  journal={Statistical science},
  pages={163--185},
  year={1998},
  publisher={JSTOR}
}

@techreport{adusumilli2018bootstrap,
  title={Bootstrap inference for propensity score matching},
  author={Adusumilli, KARUN},
  year={2018},
  institution={Working paper}
}

@article{emerson2001selection,
  title={Selection of evolutionary models for phylogenetic hypothesis testing using parametric methods},
  author={Emerson, BC and Ibrahim, KM and Hewitt, GM},
  journal={Journal of Evolutionary Biology},
  volume={14},
  number={4},
  pages={620--631},
  year={2001},
  publisher={Blackwell Publishing Ltd Oxford, UK}
}

\appendix

\section{Theoretical guarantees for sampling the copies}\label{app:sampling_copies}
\subsection{Dependence among the copies}\label{app:sampling_copies dependence}
In this section, we return to the question raised in Section~\ref{sec: sampling the copies}: since sampling $\widetilde{X}^{(1)},\dots,\widetilde{X}^{(M)}$ i.i.d.\ from the distribution $g_\pi(\,\cdot \mid  \widehat{\theta}_{1:B})$ is often computationally infeasible, can we relax this sampling step without losing finite-sample type-I error control?

In Algorithm~\ref{alg: acss-b_iid_sampling}, we assume that, after observing the posterior draws $\widehat\theta_1,\dots,\widehat\theta_B$, the copies $\widetilde{X}^{(m)}$ are then sampled i.i.d.\ from the distribution $g_\pi(\,\cdot \mid  \widehat{\theta}_{1:B})$. In practice, sampling from a complex distribution is often carried out via MCMC based strategies, and the resulting samples are only approximately i.i.d.---specifically, samples obtained by running a Markov chain will have dependence. While it is common in many sampling problems to assume mixing conditions for the Markov chain, in order to ensure that the resulting samples are approximately i.i.d., here we will use a different approach in order to ensure finite-sample validity.

Formally, define the joint sampling distribution of the copies, conditional on the data $X$ and the posterior draws $\widehat\theta_1,\dots,\widehat\theta_B$, as 
\[(\widetilde{X}^{(1)}, \dots, \widetilde{X}^{(M)}) \mid X, \widehat{\theta}_{1:B} \sim \widetilde{Q}^{M}_\pi(\,\cdot \mid X, \widehat{\theta}_{1:B}).\]
We will assume the following property of this joint distribution:
\begin{multline}\label{eq: exchangeability requirement}
    \textnormal{For any }\theta_1, \dots, \theta_B \in \Theta, \\
    \textnormal{ if }X\sim g_{\pi} \left(\,\cdot \mid \theta_{1:B}\right) \textnormal{ and } (\widetilde{X}^{(1)}, \dots, \widetilde{X}^{(M)}) \mid X \ \sim \ \widetilde{Q}^{M}_\pi(\,\cdot \mid X, \theta_{1:B}),  \\
    \textnormal{then the random vector } (X, \widetilde{X}^{(1)}, \dots, \widetilde{X}^{(M)}) \textnormal{ is exchangeable.} 
\end{multline}
For example, the i.i.d.\ sampling strategy of Algorithm~\eqref{alg: acss-b_iid_sampling} satisfies this condition---we can define $\widetilde{Q}^{M}_\pi(\,\cdot \mid X, \theta_{1:B})$ as follows:
\begin{equation}\label{eqn:QM_iid}\widetilde{Q}^{M}_\pi(\,\cdot \mid X, \theta_{1:B}) \textnormal{ is the distribution with joint density }g_\pi(x_1 \mid  \widehat{\theta}_{1:B}) \cdot \hdots \cdot g_\pi(x_M \mid  \widehat{\theta}_{1:B}).\end{equation}
However, even in settings where i.i.d.\ sampling is infeasible, we can nonetheless use MCMC strategies---e.g., the permuted serial sampler~\citep{besag1989generalized}---to draw the copies from a joint distribution $\widetilde{Q}^{M}_\pi(\,\cdot \mid X, \theta_{1:B})$ that exactly satisfies this condition (see \citet[Section 2.2.3]{acss} for more details on how to implement this in the setting of aCSS).

The aCSS-B method, with more general sampling strategy, is presented in Algorithm~\ref{alg: acss-b_general_sampling}. Of course, the original version of the method, given in Algorithm~\ref{alg: acss-b_iid_sampling}, is simply a special case obtained by choosing the i.i.d.\ sampling strategy as in~\eqref{eqn:QM_iid}.

\begin{algorithm}[ht!]
\caption{aCSS-B method (general case)}\label{alg: acss-b_general_sampling}
\KwGiven{Prior density $\pi$ on $\Theta$, and test statistic $T:\mathcal{X}\to\mathbb{R}$.}
\KwData{$X \sim f_{\theta_0}$.}

Generate $B$ posterior samples,
    \[\widehat{\theta}_1, \ldots, \widehat{\theta}_B \,\mid\, X \  \overset{\textnormal{i.i.d.}}{\sim} \ \pi(\,\cdot \mid X).\]

Generate $M$ copies of the data,
    \[\big(\widetilde{X}^{(1)}, \dots, \widetilde{X}^{(M)}\big) \,  \mid \, X, \widehat{\theta}_{1:B} 
 \ \sim \widetilde{Q}^{M}_\pi\left(\,\cdot \mid X, \widehat{\theta}_{1:B}\right),\]
 where the distribution $\widetilde{Q}^{M}_\pi(\,\cdot \mid X, \widehat{\theta}_{1:B})$ is chosen to satisfy~\eqref{eq: exchangeability requirement}.

Compute the p-value 
    \[\textnormal{pval} = \frac{1 + \sum_{m=1}^M \mathbbm{1}\{T(\widetilde{X}^{(m)}) \geq T(X)\}}{M + 1}.\]
\end{algorithm}

Our next result, proved in Appendix~\ref{app: proof of thm: validity general}, is a generalization of theorem~\ref{thm: validity}, showing that as long as the copies are sampled from a distribution satisfying~\eqref{eq: exchangeability requirement}, the same bound on type-I error still holds.
\begin{Theorem}\label{thm: validity general}
   After observing the data $X$, let $\widetilde{X}^{(1)}, \ldots, \widetilde{X}^{(M)}$ be sampled as in Algorithm~\ref{alg: acss-b_general_sampling}, for some positive prior density $\pi$ on $\Theta$. Then, if $X \sim f_{\theta_0}$ for some $\theta_0 \in \Theta$,

$$
\textnormal{d}_{\textnormal{exch}}\left(X, \widetilde{X}^{(1)}, \ldots, \widetilde{X}^{(M)}\right) \leq \inf_{\pi_0}\left\{\epsilon(\pi_0)+ \frac{\Delta(\pi_0)}{2\sqrt{B}}
\right\},$$
where the infimum is taken over all densities $\pi_0$ on $\Theta$ with respect to base measure $\nu_\Theta$ such that the support of $\bar{f}_{\pi_0}(x)$ contains the support of $f_\theta(x)$ for all $\theta$. 
\end{Theorem}

\subsection{Estimating the distribution for the copies}\label{app:sampling_copies marginal approx}
Thus far, we have considered the setting where the density $g_\pi(\,\cdot \mid  \widehat{\theta}_{1:B}) $ is computable exactly, even though drawing copies independently from this distribution may not be feasible. Next, we turn to the problem of computing this distribution itself. Recall that the density $g_\pi(\,\cdot \mid  \widehat{\theta}_{1:B}) $ is defined as
\[\propto \frac{\prod_{b=1}^B f_{\widehat\theta_b}(x)}{\bar{f}_\pi(x)^{B-1}},\]
where
\[\bar{f}_\pi(x) = \int_\Theta f_\theta(x) \pi(\theta)\;\mathsf{d}\nu_\Theta(\theta) \]
is the marginal density of $X$, integrated over the prior $\theta\sim \pi$. In many settings, evaluating the likelihood $f_\theta(x)$ at a single value $\theta$ is straightforward, but calculating the marginal likelihood can only be carried out approximately:
the integral defining the marginal likelihood is rarely available in closed form except for simple conjugate models, and is well known to be intractable in practice in many settings~\citep{chib1995marginal, gelman1995bayesian}. A variety of numerical methods have been proposed to estimate $\bar{f}_\pi(x)$, such as the Laplace approximation~\citep{barber2016Laplace}, importance sampling~\citep{newton1994approximate}, Chib’s MCMC estimator~\citep{chib1995marginal,chib2001marginal}, bridge sampling, and path sampling~\citep{meng1996simulating, gelman1998simulating}. These techniques can be leveraged to provide an estimated marginal density, $\widehat{f}_\pi(x)$, that we can then use in place of $\bar{f}_\pi(x)$ in the aCSS-B method. 
(Note that $\widehat{f}_\pi(x)$ is treated as fixed---that is, we assume this estimate of the marginal distribution is computed independently of the data used in the aCSS-B procedure.)

Specifically, we can run Algorithm~\ref{alg: acss-b_iid_sampling} with density $\widehat{g}_\pi(\,\cdot\mid \widehat\theta_{1:B})$, defined as
\[\widehat{g}_\pi(x \mid \widehat\theta_{1:B})\propto \frac{\prod_{b=1}^B f_{\widehat\theta_b}(x)}{\widehat{f}_\pi(x)^{B-1}},\]
in place of $g_\pi(\,\cdot\mid \widehat\theta_{1:B})$. (As for $\bar{f}_\pi$ earlier, now we will need to assume positivity of $\widehat{f}_\pi(x)$ in order for this to be well-defined---that is, the support of $f_\theta(x)$ is contained in the support of $\widehat{f}_\pi(x)$, for every $\theta$.) Or, more generally, if sampling i.i.d.\ copies is not feasible then we can run Algorithm~\ref{alg: acss-b_general_sampling} with a joint distribution $\widehat{Q}^M_\pi(\,\cdot\mid \widehat\theta_{1:B})$, satisfying
\begin{multline}\label{eq: exchangeability requirement (approx)}
    \textnormal{For any }\theta_1, \dots, \theta_B \in \Theta, \\
    \textnormal{ if }X\sim \widehat{g}_{\pi} \left(\,\cdot \mid \theta_{1:B}\right) \textnormal{ and } (\widetilde{X}^{(1)}, \dots, \widetilde{X}^{(M)}) \mid X \ \sim \ \widehat{Q}^{M}_\pi(\,\cdot \mid X, \theta_{1:B}),  \\
    \textnormal{then the random vector } (X, \widetilde{X}^{(1)}, \dots, \widetilde{X}^{(M)}) \textnormal{ is exchangeable.} 
\end{multline}

Of course, this modification comes at the potential cost of additional type-I error inflation, since we are now sampling the copies from an approximation to the original distribution.
The following theorem, proved in Appendix~\ref{app: proof of thm: validity general (approx)}, provides a bound on the type-I error inflation of aCSS-B, when we use an estimate $\widehat{f}_\pi$ in place of the exact marginal likelihood $\bar{f}_\pi$.

\begin{Theorem}\label{thm: validity general (approx)}
After observing the data $X$, let $\widetilde{X}^{(1)}, \ldots, \widetilde{X}^{(M)}$ be sampled via Algorithm \ref{alg: acss-b_general_sampling} except implemented with a joint distribution $\widehat{Q}^{M}_\pi(\,\cdot \mid X, \theta_{1:B})$ satisfying~\eqref{eq: exchangeability requirement (approx)}, for some positive prior density $\pi$ on $\Theta$. Assume that
    \[ \mathbb{E}_{\theta_0}\left[ \left|\left(\frac{\bar{f}_\pi(X)}{\widehat{f}_\pi(X)}\right)^{B-1} - 1\right|\right] \leq \delta_B. \]
Then, if $X \sim f_{\theta_0}$ for some $\theta_0 \in \Theta$,

$$
\textnormal{d}_{\textnormal{exch}}\left(X, \widetilde{X}^{(1)}, \ldots, \widetilde{X}^{(M)}\right) \leq \inf_{\pi_0}\left\{\epsilon(\pi_0)+ \frac{\Delta(\pi_0)}{2\sqrt{B}}
\right\} + \delta_B,$$
where the infimum is taken over all densities $\pi_0$ on $\Theta$ with respect to base measure $\nu_\Theta$ such that the support of $\bar{f}_{\pi_0}(x)$ contains the support of $f_\theta(x)$ for all $\theta$.
\end{Theorem} 
\noindent In other words, the additional term in the bound, $\delta_B$, is small whenever the estimated marginal density $\widehat{f}_\pi(x)$ is a sufficiently accurate approximation to $\bar{f}_\pi(x)$. (Note that the dependence on $B$ implies that we require a more accurate approximation $\widehat{f}_\pi$ when $B$ is large.)

\section{Proofs}\label{app: proofs}
\subsection{Proof of \texorpdfstring{\Cref{prop: posterior is sufficient}}{Proposition}}\label{app: proof of prop: posterior is sufficient}
    Since $\pi$ is assumed to be a positive density, note that
    $f_\theta(x) = \frac{\pi(\theta\mid x)}{\pi(\theta)} \cdot \bar{f}_\pi(x)$,
    where on the right-hand side, the first term depends on $x$ only via the statistic $\pi(\,\cdot\mid x)$, and the second term depends on $x$ only and not on $\theta$. By the Fisher--Neyman factorization theorem \citep[pg.~276]{casella2002statistical}, this implies that $\pi(\,\cdot\mid X)$ is a sufficient statistic of $X$. On the other hand, it is well known that the likelihood function $\theta\mapsto f_\theta(X)$ is a minimal sufficient statistic (see, e.g., \citet{fraser1963sufficiency}). Since the posterior density function $\theta\mapsto\pi(\theta\mid X)$ depends on $X$ only through the likelihood function $\theta\mapsto f_\theta(X)$, this means that the posterior is minimal sufficient.

\subsection{Proof of \texorpdfstring{\Cref{thm: validity}}{Theorem}}\label{app: proof of thm: validity}
As explained in \Cref{app:sampling_copies dependence}, by defining $\widetilde{Q}^{M}_\pi(\,\cdot \mid X, \theta_{1:B})$ as in~\eqref{eqn:QM_iid}, the result of \Cref{thm: validity} is simply a special case of \Cref{thm: validity general}---see \Cref{app: proof of thm: validity general} for the proof of this more general theorem.

\subsection{Proof of \texorpdfstring{\Cref{thm: validity general}}{Theorem}}\label{app: proof of thm: validity general}

Let $P_0$ denote the joint distribution of $(X,\widehat\theta_1,\dots,\widehat\theta_B)$, which is given by
\begin{equation}\label{eqn:proof_joint_P0}\begin{cases}
    X \sim f_{\theta_0},\\
    \{\widehat\theta_b\}_{b=1,\dots,B}\mid X \ \stackrel{\textnormal{iid}}{\sim} \ \pi(\,\cdot\mid X).
\end{cases}\end{equation}
This distribution has the following joint density at $(x,\theta_1,\dots,\theta_B)$:
\[ f_{\theta_0}(x)\cdot \prod_{b=1}^B \frac{f_{\theta_b}(x)\pi(\theta_b)}{\bar{f}_\pi(x)}\]
(with respect to the base measure $\nu_{\mathcal{X}}\times\nu_\Theta\times\dots\times\nu_\Theta$).
 The aCSS-B procedure can be equivalently characterized as
\begin{equation}\label{eqn:proof_joint_P0_with_copies}\begin{cases}
    (X,\widehat\theta_1,\dots,\widehat\theta_B) \sim P_0, \\
    (\widetilde{X}^{(1)},\dots,\widetilde{X}^{(M)})\mid (X,\widehat\theta_1,\dots,\widehat\theta_B) \sim \widetilde{Q}_\pi^M(\,\cdot\mid X,\widehat{\theta}_{1:B}).
\end{cases}\end{equation}
Our goal, then, is to bound the distance to exchangeability induced by this distribution.

\emph{Step 1: Defining $P_1$ to bound distance to exchangeability.} We next define a joint distribution $P_1$, with joint density at $(x,\theta_1,\dots,\theta_B)$ given by:
\[ \frac{1}{B}\sum_{b=1}^B \frac{\pi_0(\theta_b)}{\pi(\theta_b)} \cdot \bar{f}_\pi(x)\cdot \prod_{b=1}^B \frac{f_{\theta_b}(x)\pi(\theta_b)}{\bar{f}_\pi(x)}  . \]
(We can verify that this expression integrates to $1$ by definition of $\bar{f}_\pi$, and therefore defines a valid joint density.)
Note that, by construction, under the joint distribution $P_1$ we have
\[X\mid (\widehat\theta_1,\dots,\widehat\theta_B) \sim g_\pi(\,\cdot\mid\widehat{\theta}_{1:B}),\]
where we recall that $g_\pi(\,\cdot\mid\widehat{\theta}_{1:B})$ is the distribution with density $\propto \frac{\prod_{b=1}^B f_{\widehat\theta_b}(x)}{\bar{f}_\pi(x)^{B-1}}$. Therefore, if we consider the sampling distribution
\begin{equation}\label{eqn:proof_joint_P1_with_copies}\begin{cases}
    (X,\widehat\theta_1,\dots,\widehat\theta_B) \sim P_1, \\
    (\widetilde{X}^{(1)},\dots,\widetilde{X}^{(M)})\mid (X,\widehat\theta_1,\dots,\widehat\theta_B) \sim \widetilde{Q}_\pi^M(\,\cdot\mid X,\widehat{\theta}_{1:B}),
\end{cases}\end{equation}
then by~\eqref{eq: exchangeability requirement} on $\widetilde{Q}_\pi^M(\,\cdot\mid X,\widehat{\theta}_{1:B})$, it holds that $(X,\widetilde{X}^{(1)},\dots,\widetilde{X}^{(M)})$ are exchangeable conditional on $(\widehat\theta_1,\dots,\widehat\theta_B)$, and consequently are also exchangeable after marginalizing over $(\widehat\theta_1,\dots,\widehat\theta_B)$.

Consequently, we can see that $ \textnormal{d}_{\textnormal{exch}}(X, \widetilde{X}^{(1)}, \ldots, \widetilde{X}^{(M)}) $ is bounded by the total variation distance between the sampling distributions given in~\eqref{eqn:proof_joint_P0_with_copies} and in~\eqref{eqn:proof_joint_P1_with_copies}, which can be simplified to
\[ \textnormal{d}_{\textnormal{exch}}\left(X, \widetilde{X}^{(1)}, \ldots, \widetilde{X}^{(M)}\right) \leq \textnormal{d}_{\textnormal{TV}}(P_0,P_1),\]
since the distributions~\eqref{eqn:proof_joint_P0_with_copies} and~\eqref{eqn:proof_joint_P1_with_copies} differ only in their first line. From this point on, then, we only need to bound this last total variation distance.

\emph{Step 2: Defining $P_{0.5}$.}\
First, we define an intermediate distribution, $P_{0.5}$, with joint density at $(x,\theta_1,\dots,\theta_B)$ given by
\[ \frac{1}{B}\sum_{b=1}^B \frac{\pi_0(\theta_b)/\bar{f}_{\pi_0}(x)}{\pi(\theta_b)/\bar{f}_\pi(x)} \cdot f_{\theta_0}(x)\cdot \prod_{b=1}^B \frac{f_{\theta_b}(x)\pi(\theta_b)}{\bar{f}_\pi(x)},\]
where
\[\bar{f}_{\pi_0}(x) = \int_\Theta f_\theta(x)\pi_0(\theta)\;\mathsf{d}\nu_\Theta(\theta)\]
is the marginal likelihood corresponding to the prior $\pi_0$. We can then write 
\[\textnormal{d}_{\textnormal{TV}}(P_0,P_1) \leq 
\textnormal{d}_{\textnormal{TV}}(P_0,P_{0.5}) + \textnormal{d}_{\textnormal{TV}}(P_{0.5},P_1) .\]
We now bound the remaining two terms separately.

\emph{Step 3: Bounding $\textnormal{d}_{\textnormal{TV}}(P_{0.5},P_1)$.}\ Note that we can express the distribution $P_{0.5}$ as a mixture, $P_{0.5} = \frac{1}{B}\sum_{b=1}^B P_{0.5,b}$, where $P_{0.5,b}$ has joint density
\[f_{\theta_0}(x)\cdot \frac{f_{\theta_b}(x)\pi_0(\theta_b)}{\bar{f}_{\pi_0}(x)} \cdot \prod_{b'\neq b}\frac{f_{\theta_{b'}}(x)\pi(\theta_{b'})}{\bar{f}_\pi(x)},\]
which can be interpreted as follows: 
\begin{equation}\label{eqn:proof_joint_P0.5b}\begin{cases}
    X \sim f_{\theta_0},\\
    \widehat\theta_b \mid X \ \sim \ \pi_0(\,\cdot\mid X),\\
    \{\widehat\theta_{b'}\}_{b'\neq b}\mid X,\widehat\theta_b \ \stackrel{\textnormal{iid}}{\sim} \ \pi(\,\cdot\mid X).
\end{cases}\end{equation}
Similarly, we can express the distribution $P_1$ as a mixture, $P_1 = \frac{1}{B}\sum_{b=1}^B P_{1,b}$, where $P_{1,b}$ has joint density
\[ \pi_0(\theta_b) \cdot f_{\theta_b}(x) \cdot \prod_{b'\neq b}\frac{f_{\theta_{b'}}(x)\pi(\theta_{b'})}{\bar{f}_\pi(x)}  ,\]
which can be interpreted as follows: 
\[\begin{cases}
    \widehat\theta_b\sim \pi_0,\\
    X \mid \widehat\theta_b  \ \sim \ f_{\widehat\theta_b},\\
    \{\widehat\theta_{b'}\}_{b'\neq b}\mid X,\widehat\theta_b \ \stackrel{\textnormal{iid}}{\sim} \ \pi(\,\cdot\mid X).
\end{cases}\]
Equivalently, by swapping the order in which we draw $\widehat\theta_b$ and $X$, this is the same as
\begin{equation}\label{eqn:proof_joint_P1b}\begin{cases}
    X \sim \bar{f}_{\pi_0},\\
    \widehat\theta_b\mid X \sim \pi_0(\,\cdot\mid X),\\
    \{\widehat\theta_{b'}\}_{b'\neq b}\mid X,\widehat\theta_b \ \stackrel{\textnormal{iid}}{\sim} \ \pi(\,\cdot\mid X).
\end{cases}\end{equation}
By comparing~\eqref{eqn:proof_joint_P0.5b} and~\eqref{eqn:proof_joint_P1b}, we can see that
\[\textnormal{d}_{\textnormal{TV}}(P_{0.5,b},P_{1,b})  = \textnormal{d}_{\textnormal{TV}}(f_{\theta_0},\bar{f}_{\pi_0}),\]
and moreover, $\textnormal{d}_{\textnormal{TV}}(f_{\theta_0},\bar{f}_{\pi_0}) \leq \epsilon(\pi_0)$ by our definition of $\epsilon(\pi_0)$ (see \Cref{def:definition_1}). Then
\[\textnormal{d}_{\textnormal{TV}}(P_{0.5},P_1) 
= \textnormal{d}_{\textnormal{TV}}\left(\frac{1}{B}\sum_{b=1}^B P_{0.5,b} , \frac{1}{B}\sum_{b=1}^B P_{1,b}\right) \leq  \frac{1}{B}\sum_{b=1}^B \textnormal{d}_{\textnormal{TV}}(P_{0.5,b},P_{1,b}) \leq \epsilon(\pi_0). \]

\emph{Step 4: Bounding $\textnormal{d}_{\textnormal{TV}}(P_0,P_{0.5})$.}\
By comparing~\eqref{eqn:proof_joint_P0} with~\eqref{eqn:proof_joint_P0.5b}, we can see that the marginal distribution of $X$ is the same for both (i.e., $X\sim f_{\theta_0}$), while the conditional distribution of $(\widehat\theta_1,\dots,\widehat\theta_B)\mid X$ is different: under $P_0$, these $B$ values are drawn i.i.d.\ from the posterior $\pi(\,\cdot\mid X)$, while under $P_{0.5}$, we instead draw one $\widehat\theta_b$ (with $b$ selected uniformly at random) from $\pi_0(\,\cdot\mid X)$, and the remaining values $\{\widehat\theta_{b'}\}_{b'\neq b}$ are drawn i.i.d.\ from $\pi(\,\cdot\mid X)$. We will now need the following lemma:
\begin{Lemma}\label{lemma:TV_bound}
Let $P$ and $Q$ be probability measures on a measurable space $(\Omega,\mathcal F)$ with $Q\ll P$ and let $B>1$ be an integer. Let $Z=\left(Z_1, \ldots, Z_B\right)$, where $Z_1, \ldots, Z_B \overset{\text { iid }}{\sim} P$. Define a corrupted vector $Z^{\prime}=\left(Z_1^{\prime}, \ldots, Z_B^{\prime}\right)$, where we draw a random $K \sim$ Uniform $\{1, \ldots, B\}$, and sample $Z_K^{\prime} \sim Q$, and set $Z_k^{\prime}=Z_k$ for all $k \neq K$. Then
$$
\textnormal{d}_{\textnormal{TV}}\left(Z, Z^{\prime}\right) \leq \frac{1}{2}\sqrt{\frac{\textnormal{d}_{\chi^2}(Q \| P)}{B}} .
$$ 
\end{Lemma}
Applying this Lemma, we then have
\[\textnormal{d}_{\textnormal{TV}}\left( (P_0)_{\widehat\theta_{1:B}\mid X} , (P_{0.5})_{\widehat\theta_{1:B}\mid X} \right)\leq \frac{1}{2}\sqrt{\frac{\textnormal{d}_{\chi^2}\big(\pi_0(\,\cdot \mid X) \,\|\, \pi(\,\cdot\mid X)\big)}{B}},\]
where $(P_0)_{\widehat\theta_{1:B}\mid X}$ denotes the conditional distribution of $(\widehat\theta_1,\dots,\widehat\theta_B)\mid X$ under the joint distribution $P_0$, and same for $P_{0.5}$. Therefore,
\[\textnormal{d}_{\textnormal{TV}}(P_0,P_{0.5})= \mathbb{E}\left[\textnormal{d}_{\textnormal{TV}}\left( (P_0)_{\widehat\theta_{1:B}\mid X} , (P_{0.5})_{\widehat\theta_{1:B}\mid X} \right)\right]\leq \frac{\Delta(\pi_0)}{2\sqrt{B}},\]
by definition of $\Delta(\pi_0)$ (see \Cref{def:definition_2}).

\subsection{Proof of \texorpdfstring{\Cref{lemma:TV_bound}}{Lemma}}
By construction, the distribution of $Z'$ can be written as
$$
\widetilde Q
\;=\;
\frac1B\sum_{b=1}^B
\left(P^{b-1}\otimes Q\otimes P^{B-b}\right)
\;\ll\;P^B.
$$
If we denote $f = \frac{\mathsf{d}Q}{\mathsf{d}P}$, then for each $b$,
$$
\frac{\mathsf{d}\,(P^{b-1}\times Q\times P^{B-b})}{\mathsf{d}P^B}(x_1,\dots,x_B)
= f(x_b),
$$
for $P^B$-almost-every $(x_1,\dots,x_B)$.
Hence we can calculate the Radon–Nikodym derivative
$$
\frac{\mathsf{d}\widetilde Q}{\mathsf{d}P^B}(x_1,\dots,x_B)
= \frac1B\sum_{b=1}^B f(x_b),
$$
for $P^B$-almost-every $(x_1,\dots,x_B)$.
By definition of total variation,
$$
\textnormal{d}_{\textnormal{TV}}(P^B,\widetilde Q)
= \frac12
\int_{\Omega^B}
  \left|\,1 - \frac1B\sum_{b=1}^B f(x_b)\right|
\,\mathsf{d}P^B(x).
$$
Now define $\Delta(x)=1-f(x)$.
Note that
\[\int_\Omega \Delta(x)\;\mathsf{d}P(x) = \int_\Omega (1-f(x))\;\mathsf{d}P(x)=0\]
and
\[\int_\Omega \Delta(x)^2\;\mathsf{d}P(x) = \int_\Omega (1-f(x))^2\;\mathsf{d}P(x)=\textnormal{d}_{\chi^2}(Q\| P),\]
by definition of $f$. 
Then
\begin{multline*}
\textnormal{d}_{\textnormal{TV}}(P^B,\widetilde Q)
= \frac1{2B}
\int_{\Omega^B}
  \left|\sum_{b=1}^B \Delta(x_b)\right|
\,\mathsf{d}P^B(x)
\leq \frac1{2B}
\left[\int_{\Omega^B}
  \left(\sum_{b=1}^B \Delta(x_b)\right)^2
\,\mathsf{d}P^B(x) \right]^{1/2}\\= \frac{1}{2B}\left[\sum_{b=1}^B\sum_{b'=1}^B \int_{\Omega^B} \Delta(x_b)\Delta(x_{b'})\;\mathsf{d}P^B(x)\right]^{1/2},\end{multline*}
by Cauchy--Schwarz. But for $b\neq b'$, we have
\[\int_{\Omega^B} \Delta(x_b)\Delta(x_{b'})\;\mathsf{d}P^B(x) = \left(\int_\Omega \Delta(x_b)\;\mathsf{d}P(x_b)\right) \cdot \left(\int_\Omega \Delta(x_{b'})\;\mathsf{d}P(x_{b'})\right) = 0,\]
while for the case $b=b'$ we have
\[\int_{\Omega^B} \Delta(x_b)^2\;\mathsf{d}P^B(x) = \textnormal{d}_{\chi^2}(Q\|P).\]
Therefore,
\[\textnormal{d}_{\textnormal{TV}}(P^B,\widetilde Q) \leq  \frac{1}{2B}\left[\sum_{b=1}^B \sum_{b'=1}^B \textnormal{d}_{\chi^2}(Q\|P)\cdot \mathbbm{1}_{b=b'} \right]^{1/2}= \frac{\sqrt{B\textnormal{d}_{\chi^2}(Q\|P)}}{2B},\]
which completes the proof.

\subsection{Proof of \texorpdfstring{\Cref{thm: validity general (approx)}}{Theorem}}\label{app: proof of thm: validity general (approx)}

Recall from the proof of \Cref{thm: validity general}, the joint distribution $P_1$ on $(X,\widehat\theta_1,\dots,\widehat\theta_B)$, which we defined to have the joint density
\[ \frac{1}{B}\sum_{b=1}^B \frac{\pi_0(\theta_b)}{\pi(\theta_b)} \cdot \bar{f}_\pi(x)\cdot \prod_{b=1}^B \frac{f_{\theta_b}(x)\pi(\theta_b)}{\bar{f}_\pi(x)}  . \]
Now define a distribution $P_2$, with joint density
\[ \frac{\left(\frac{\bar{f}_\pi(x)}{\widehat{f}_\pi(x)}\right)^{B-1}}{\mathbb{E}_{\bar{f}_{\pi_0}}\left[\left(\frac{\bar{f}_\pi(X)}{\widehat{f}_\pi(X)}\right)^{B-1}\right]} \cdot \frac{1}{B}\sum_{b=1}^B \frac{\pi_0(\theta_b)}{\pi(\theta_b)} \cdot \bar{f}_\pi(x)\cdot \prod_{b=1}^B \frac{f_{\theta_b}(x)\pi(\theta_b)}{\bar{f}_\pi(x)}  , \]
which differs from $P_1$ only in the presence of the first term. We can observe that, under $P_2$, the conditional distribution of $X\mid(\widehat\theta_1,\dots,\widehat\theta_B)$ is given by
\[X\mid (\widehat\theta_1,\dots,\widehat\theta_B) \sim \widehat{g}_\pi(\,\cdot\mid\widehat{\theta}_{1:B}),\]
where we recall that this joint density is defined as $\widehat{g}_\pi(x\mid\widehat{\theta}_{1:B})\propto \frac{\prod_{b=1}^B f_{\widehat\theta_b}(x)}{\widehat{f}_\pi(x)^{B-1}}$. Therefore, if we consider the sampling distribution
\begin{equation}\label{eqn:proof_joint_P2_with_copies}\begin{cases}
    (X,\widehat\theta_1,\dots,\widehat\theta_B) \sim P_2, \\
    (\widetilde{X}^{(1)},\dots,\widetilde{X}^{(M)})\mid (X,\widehat\theta_1,\dots,\widehat\theta_B) \sim \widehat{Q}_\pi^M(\,\cdot\mid X,\widehat{\theta}_{1:B}),
\end{cases}\end{equation}
then by~\eqref{eq: exchangeability requirement (approx)} on $\widehat{Q}_\pi^M(\,\cdot\mid X,\widehat{\theta}_{1:B})$, it holds that $(X,\widetilde{X}^{(1)},\dots,\widetilde{X}^{(M)})$ are exchangeable conditional on $(\widehat\theta_1,\dots,\widehat\theta_B)$, and consequently are also exchangeable after marginalizing over $(\widehat\theta_1,\dots,\widehat\theta_B)$.

Now compare this to running aCSS-B with the approximated marginal, which can be characterized by the sampling distribution
\[\begin{cases}
    (X,\widehat\theta_1,\dots,\widehat\theta_B) \sim P_0, \\
    (\widetilde{X}^{(1)},\dots,\widetilde{X}^{(M)})\mid (X,\widehat\theta_1,\dots,\widehat\theta_B) \sim \widehat{Q}_\pi^M(\,\cdot\mid X,\widehat{\theta}_{1:B}).
\end{cases}\]
Comparing this with~\eqref{eqn:proof_joint_P2_with_copies}, we can see that
\[ \textnormal{d}_{\textnormal{exch}}\left(X, \widetilde{X}^{(1)}, \ldots, \widetilde{X}^{(M)}\right) \leq \textnormal{d}_{\textnormal{TV}}(P_0,P_2),\]
similarly to the proof of \Cref{app: proof of thm: validity general}. Next, we have already shown in the proof of \Cref{app: proof of thm: validity general} that $\textnormal{d}_{\textnormal{TV}}(P_0,P_1)\leq \epsilon(\pi_0) + \frac{\Delta(\pi_0)}{2\sqrt{B}}$, so from this point on we only need to bound $\textnormal{d}_{\textnormal{TV}}(P_1,P_2)$. 

By definition of the joint density for $P_2$ as compared to $P_1$, we can compute the Radon--Nikodym derivative as
\[\frac{\mathsf{d}P_2(x,\theta_1,\dots,\theta_B)}{\mathsf{d}P_1(x,\theta_1,\dots,\theta_B)}
= \frac{\left(\frac{\bar{f}_\pi(x)}{\widehat{f}_\pi(x)}\right)^{B-1}}{\mathbb{E}_{\bar{f}_{\pi_0}}\left[\left(\frac{\bar{f}_\pi(X)}{\widehat{f}_\pi(X)}\right)^{B-1}\right]}.\]
If the denominator is $\leq 1$, then we have
\[
    \textnormal{d}_{\textnormal{TV}}(P_1,P_2) = \mathbb{E}_{P_1}\left[\left(1 - \frac{\mathsf{d}P_2(X,\widehat\theta_1,\dots,\widehat\theta_B)}{\mathsf{d}P_1(X,\widehat\theta_1,\dots,\widehat\theta_B)}\right)_+\right]
    \leq \mathbb{E}_{P_1}\left[\left(1 -\left(\frac{\bar{f}_\pi(X)}{\widehat{f}_\pi(X)}\right)^{B-1}\right)_+\right], 
\]
while if instead the denominator is $\geq 1$ then
\[
    \textnormal{d}_{\textnormal{TV}}(P_1,P_2) = \mathbb{E}_{P_1}\left[\left( \frac{\mathsf{d}P_2(X,\widehat\theta_1,\dots,\widehat\theta_B)}{\mathsf{d}P_1(X,\widehat\theta_1,\dots,\widehat\theta_B)}-1 \right)_+\right]
    \leq \mathbb{E}_{P_1}\left[\left(\left(\frac{\bar{f}_\pi(X)}{\widehat{f}_\pi(X)}\right)^{B-1}-1\right)_+\right].
\]
In either case, then,
\[
    \textnormal{d}_{\textnormal{TV}}(P_1,P_2)
    \leq \mathbb{E}_{P_1}\left[\left|\left(\frac{\bar{f}_\pi(X)}{\widehat{f}_\pi(X)}\right)^{B-1}-1\right|\right]
    = \mathbb{E}_{\theta_0}\left[\left|\left(\frac{\bar{f}_\pi(X)}{\widehat{f}_\pi(X)}\right)^{B-1}-1\right|\right],\]
where the last step holds 
since under $P_1$, the marginal distribution of $X$ is given by $f_{\theta_0}$.
This verifies that $\textnormal{d}_{\textnormal{TV}}(P_1,P_2)\leq \delta_B$, which completes the proof.

\section{Sampling details}
\label{app: sampling details}
\subsection{Logistic Regression}{\label{app:logistic_regression_computation}}

This section gives the implementation details for the logistic regression experiment presented in \Cref{sec: simulation logistic regression}. (For the comparison to aCSS, the implementation of aCSS is exactly as in \citet[Example 1, Section 4.5.2]{acss}.)

\emph{Sampling from the posterior:}\ 
The posterior distribution is challenging to sample from directly, so we instead sample the draws $\widehat\theta_b$ using Metropolis--Hastings sampling.

We now give details.
The (unnormalized) log-density of the posterior distribution is given by
$$
\Psi(\theta;x)=\log f_\theta(x)+\log \pi(\theta),
$$
where 
\begin{align*}
    f_{\theta}(x)&=\prod_{i=1}^n\left(\frac{e^{Z_i^{\top} \theta}}{1+e^{Z_i^{\top} \theta}}\right)^{x_i} \cdot\left(\frac{1}{1+e^{Z_i^{\top} \theta}}\right)^{1-x_i},\\
    \pi(\theta) &= \prod_{j=1}^d \phi\left(\theta_j; 0, 1\right).
\end{align*}
We will use a Laplace approximation to the posterior distribution as the proposal distribution. Define
$$
\widehat{\theta}=\arg\max_\theta \Psi(\theta) .
$$
Since this does not admit a closed-form solution, in practice we solve this optimization problem numerically via \texttt{optim} in \texttt{R}. By a second-order Taylor expansion to $\Psi$ around $\widehat{\theta}$, we have
$$
\Psi(\theta) \approx \Psi_{\textnormal{Laplace}}(\theta):=\Psi(\widehat{\theta})-\frac{1}{2}(\theta-\widehat{\theta})^{\top} \mathbb{H}(\theta-\widehat{\theta}),
$$
where $\mathbb{H}$ is the Hessian of the negative log posterior evaluated at $\widehat{\theta}$ :
$$
\mathbb{H}=-\left.\nabla^2 \Psi(\theta)\right|_{\theta=\widehat{\theta}} = \sum_{i=1}^n \frac{e^{Z_i^\top
\widehat\theta}}{\left(1+e^{Z_i^\top\widehat\theta}\right)^2}\,Z_i\,Z_i^\top
\;+\;\mathbb{I}_d.
$$
Therefore, $\Psi(\theta)$ is approximately equal to $\Psi_{\textnormal{Laplace}}(\theta)$, which is the (unnormalized) log-density of a multivariate Gaussian distribution, $\mathcal{N}(\widehat\theta,\mathbb{H}^{-1})$.
To sample the posterior draws $\widehat\theta_b$, we therefore run the Metropolis--Hastings algorithm \citep{owen_mcbook}, with proposal distribution $\mathcal{N}(\widehat\theta,\mathbb{H}^{-1})$. We use a burn-in of $500$ steps, and then extract samples $\widehat\theta_1,\dots,\widehat\theta_B$ at every tenth step.

\emph{Sampling the copies:}\ Next we give details on sampling the copies $\widetilde{X}^{(m)}$. Recall that our goal is to sample the copies i.i.d.\ from the distribution with density defined as in~\eqref{eqn:conditional_density_X_approx}. We will first compute an approximation to the marginal density (recall \Cref{app:sampling_copies marginal approx}). Using the same Laplace approximation as above, we replace $\bar{f}_\pi(x)$ with 
\begin{multline*}\widehat{f}_\pi(x) \propto  \int \exp (\Psi_{\textnormal{Laplace}}(\theta)) \;\textnormal{d}\theta = \int \exp (\Psi(\widehat{\theta})) \exp \left(-\frac{1}{2}(\theta-\widehat{\theta})^{\top} \mathbb{H}(\theta-\widehat{\theta})\right) \textnormal{d} \theta\\
    =\exp (\Psi(\widehat{\theta})) \sqrt{\frac{(2 \pi)^d}{\operatorname{det} (\mathbb{H})}}.\end{multline*}
(Note that this right-hand side is, implicitly, a function of $x$, since $\widehat\theta$ and $\mathbb{H}$ both depend on $x$.)
This then leads to an approximate density, $\widehat{g}_\pi(x\mid \widehat\theta_1,\dots,\widehat\theta_B)$ (as in~\eqref{eqn:conditional_density_X_approx}) from which to sample the copies.
This density then serves as the target density in a Gibbs sampling algorithm \citep{owen_mcbook}. We use the permuted serial sampler~\citep{besag1989generalized}, as follows:
\begin{enumerate}
\item \emph{Initialization.} Draw $m_0\in\{0,\dots,M\}$ uniformly at random, and set
\begin{align*}
&\widetilde X^{(m_0)} \;\leftarrow\; X,
\end{align*}

\item \emph{Iterations.} For $t=m_0+1,\dots,M$, for each \(i=1,\dots,n\), 
\begin{align*}
    \text{sample }\widetilde X_i^{(t)} \sim \widehat g _{\pi}(\cdot \mid x_{-i},\widehat\theta_{1:B}) \text{ using } x_{-i} = (\widetilde X_{<i}^{(t)},\widetilde X_{>i}^{(t-1)}),
\end{align*}
where $\widehat g _{\pi}(\cdot \mid x_{-i},\widehat\theta_{1:B})$ is the conditional density induced by $\widehat{g}_\pi(x\mid \widehat\theta_1,\dots,\widehat\theta_B)$. Similarly, for $t=m_0-1,\dots,0$, for each $i=n,\dots,1$,
\begin{align*}
    \text{sample }\widetilde X_i^{(t)} \sim \widehat g _{\pi}(\cdot \mid x_{-i},\widehat\theta_{1:B}) \text{ using } x_{-i} = (\widetilde X_{<i}^{(t+1)},\widetilde X_{>i}^{(t)}).
\end{align*}
Since each $\widetilde X_{i}^{(t)}$ is Bernoulli, we can compute the probability explicitly as $$\mathbb{P}(X_i^{(t)}=0) = \frac{\widehat g _{\pi}(0 \mid x_{-i},\widehat\theta_{1:B})}{\widehat g _{\pi}(0 \mid x_{-i},\widehat\theta_{1:B}) + \widehat g _{\pi}(1 \mid x_{-i},\widehat\theta_{1:B})}$$ which allows us to draw from the conditional distribution.
\item \emph{Output.} Discard $\widetilde X^{(m_0)}$ and return copies $\widetilde{X}^{(0)},\dots,\widetilde{X}^{(m_0-1)},\widetilde{X}^{(m_0+1)}, \dots, \widetilde{X}^{(M)}$.\footnote{For the serial sampler to return exchangeable copies, as required in~\eqref{eq: exchangeability requirement} or~\eqref{eq: exchangeability requirement (approx)}, formally we would also need to randomly permute the set of copies produced by this algorithm---but since the p-value is invariant to shuffling the copies, it is unnecessary for this last step.}
\end{enumerate}

\subsection{Mixture of Gaussians}\label{app: Gaussian mixture}

This section gives the implementation details for the mixture of Gaussians experiment
presented in \Cref{sec: simulation Gaussian mixture}. (For the comparison to reg-aCSS, the implementation of reg-aCSS is exactly as in \citet[Section 6.1]{reg_acss}.)

\emph{Sampling from the posterior:}\ Define $\theta = (w_1,\mu_1,\sigma^2_1,\mu_2,\sigma^2_2)$. To sample from the posterior of $\theta$ we introduce the latent variable $Z \in \{1,2\}^n$. 
\begin{align*}
    X_i\mid Z_i=j \sim \mathcal{N}(\mu_j,\sigma_j^2), \quad \mathbb{P}(Z_i = j)=w_j,
\end{align*}
for each $j=1,2$, where for convenience we write $w_2 = 1-w_1$.
This makes sampling from the posterior of $\theta$ straightforward. The full conditional distributions are given by the following: 
\begin{align}
\label{eqn:mixGaus_Z}&\quad \mathbb{P}(Z_i = j \mid X, w_1,\mu_1,\sigma^2_1,\mu_2,\sigma^2_2)
=\frac{w_j\,\phi(x_i;\mu_j,\sigma_j^2)}
{\sum_{\ell=1}^2 w_\ell\,\phi(x_i;\mu_\ell,\sigma_\ell^2)},
\quad i=1,\dots,n,\\
\label{eqn:mixGaus_w}&\quad w_1\mid Z\;\sim\;\textnormal{Beta}\left(2+n_1,2+n_2\right),
\quad n_j=\sum_{i=1}^n\mathbf1\{Z_i=j\},\\
\label{eqn:mixGaus_mu}&\quad \mu_j\mid \sigma_j^2,Z,X
\;\sim\;
\mathcal N\!\left(\frac{\sum_{i:z_i = j}x_i}{1+n_j},\,\frac{\sigma_j^2}{1+n_j}\right),\\
\label{eqn:mixGaus_sig}&\quad \sigma_j^2\mid \mu_j,Z,X
\;\sim\;
\textnormal{Inv-Gamma}\!\left({\frac{3}{2}}+\frac{n_j}{2},\;
\frac{1}{2}+\frac{1}{2}\sum_{i:Z_i=j}(x_i-\mu_j)^2+\frac{1}{2}\mu_j^2\right).
\end{align}
These allow for efficient sampling using the Gibbs sampler. 

We begin by initializing the two–component mixture as follows:
\[
w^{(0)} = \bigl(w_1^{(0)},w_2^{(0)}\bigr)
= \Bigl(\tfrac12,\;\tfrac12\Bigr).
\]
The latent allocations are then initialized by splitting the data at its sample median:
\[
Z_i^{(0)} =
\begin{cases}
1, & X_i \le \mathrm{median}(X),\\
2, & X_i > \mathrm{median}(X),
\end{cases}
\quad i=1,\dots,n.
\]
Conditional on these initial assignments, we set each component’s parameters to the empirical moments of its cluster:
\[
\mu_j^{(0)}
= \frac{1}{n_j^{(0)}}\sum_{i:Z_i^{(0)}=j} X_i,
\qquad
\sigma_j^{2\,(0)}
= \frac{1}{n_j^{(0)}}\sum_{i:Z_i^{(0)}=j}\bigl(X_i - \mu_j^{(0)}\bigr)^2,
\quad
n_j^{(0)} = \sum_{i=1}^n \mathbf1\{Z_i^{(0)}=j\}.
\]
Thereafter, for each iteration $t$, we perform the following Gibbs‐sampling updates:
\begin{enumerate}
  \item Independently for each $i$, sample each $Z_i^{(t)}\mid X,\,w^{(t-1)},\,\mu^{(t-1)},\,\sigma^{2,(t-1)}$ according to \eqref{eqn:mixGaus_Z}.
  \item Sample 
    $
      w_1^{(t)} \,\bigm|\,Z^{(t)}$ according to~\eqref{eqn:mixGaus_w}, and set $w_2^{(t)} = 1 - w_1^{(t)}$.
  \item For each \(j=1,2\), draw
    \(\mu_j^{(t)}\mid \sigma_j^{2,(t-1)},Z^{(t)},X\)
    and
    \(\sigma_j^{2,(t)}\mid \mu_j^{(t)},Z^{(t)},X\)
    according to~\eqref{eqn:mixGaus_mu} and~\eqref{eqn:mixGaus_sig}, respectively.
\end{enumerate}

We discard the first $500$ draws and extract the posterior samples $\widehat\theta_1,\dots,\widehat\theta_B$ at every tenth step.

\emph{Sampling the copies:}\
To sample the copies, we again use an approximation of the marginal $\bar f_\pi(x)$ and sample from
\[
\widehat g_\pi(x \mid \widehat{\theta}_{1:B}) \propto 
\frac{\prod_{b=1}^B f_{\widehat{\theta}_b}(x)}{\widehat{f}_\pi(x)^{\,B-1}}.
\]
In order to approximate the marginal, we note that
\[
\bar f_\pi(x)=\int f_\theta(x)\,\pi(\theta)\,d\theta,
\]
and define the log of the unnormalized posterior as 
\[
\Psi(\theta;x)=\log f_\theta(x)+\log \pi(\theta).
\]
For our choice of priors it takes the following form: 
\[
\Psi(\theta;x) 
= \sum_{i=1}^n \log\!\left( 
    w_1 \frac{1}{\sqrt{2\pi\sigma_1^2}} 
         \exp\!\left(-\frac{(x_i-\mu_1)^2}{2\sigma_1^2}\right) 
  + (1-w_1) \frac{1}{\sqrt{2\pi\sigma_2^2}} 
         \exp\!\left(-\frac{(x_i-\mu_2)^2}{2\sigma_2^2}\right)
\right)
\]
\[
+ \log w_1 + \log(1-w_1)
+ \sum_{j=1}^2 \left[
\log(0.5) - 2\log(\sigma_j^2) - \frac{0.5}{\sigma_j^2}
- \tfrac{1}{2}\log(2\pi\sigma_j^2) - \frac{\mu_j^2}{2\sigma_j^2}
\right].
\]

In our setting since we have two components in the null mixture distribution, the posterior is invariant to the swap of the component labels and has \emph{two} modes. A single Laplace approximation is therefore inaccurate. We approximate $\bar f_\pi(x)$ by a mixture of two normal distributions where the means of the two components are identified as the two local maxima of $\Psi(\theta; x)$; say $\widetilde\theta_1$ and $\widetilde\theta_2$. Following the same idea as in the usual construction of a Laplace approximation, the variance of these individual Gaussian components would be the inverse of the determinant of the Hessian of $\Psi(\theta; x)$ at $\theta = \widetilde\theta_1$ and $\widetilde\theta_2$, respectively, where the Hessians at these two points are defined as
\begin{align*}
\mathbb{H}_1 = -\nabla^2 \Psi(\theta)\vert_{\theta = \widetilde\theta_1} \qquad\text{and} \qquad
\mathbb{H}_2 = -\nabla^2 \Psi(\theta)\vert_{\theta = \widetilde\theta_2},
\end{align*}
both computable in closed form. Our two–mode Laplace approximation to the marginal is then
\[
\widehat{f}_\pi(x)\;=\; e^{L_1}+e^{L_2},
\]
where 
\[
L_s \;=\; \Psi(\widetilde\theta_s;x) \;-\; \tfrac{1}{2}\log\!\det \mathbb{H}_s
\;+\; \tfrac{5}{2}\log(2\pi), \qquad s=1,2
\]
is the contribution to the integral from the $s^{\text{th}}$ component of the normal mixture approximation to the (unnormalized) posterior. 
In order to obtain $\widehat \theta_s$, we use \texttt{optim} in \texttt{R} as follows:
\begin{enumerate}
    \item \emph{Reparametrization.} To convert the constrained optimization into an unconstrained one, define the following reparameterization for $\theta = (w_1, \mu_1, \sigma_1^2, \mu_2, \sigma_1^2)$:
    \begin{align*}
        \vartheta(\theta) = (z_1, \mu_1, s_1, \mu_2, s_2) \text{ where } z_1 = \log\left(\frac{w_1}{1-w_1}\right), &\ s_1 = \log(\sigma_1^2),\text{ and }s_2 = \log(\sigma_2^2).
    \end{align*}
    \item \emph{Initialization.} Apply k-means with $k=2$ on the observed data and obtain two clusters $\mathcal{C}_1$ and $\mathcal{C}_2$ such that $\mathcal{C}_1 \cup \mathcal{C}_2 = \{1,\dots,n\}$  with the cluster centers $c_1$ and $c_2$ respectively. Define $\tau_1^2 = \frac{1}{|\mathcal{C}_1|-1} \sum_{i \in \mathcal{C}_1} (x_i - c_1)^2$ and $\tau_2^2 = \frac{1}{|\mathcal{C}_2|-1} \sum_{i \in \mathcal{C}_2} (x_i - c_2)^2$. Define the following two initializations:
    \begin{itemize}
        \item $\vartheta_1^{(0)} = \vartheta\left(\frac{|\mathcal{C}_1|}{n}, c_1, \tau_1^2, c_2, \tau_2^2\right)$,
        \item $\vartheta_2^{(0)} = \vartheta\left(\frac{|\mathcal{C}_2|}{n}, c_2, \tau_2^2, c_1, \tau_1^2\right)$.
    \end{itemize}
    \item \emph{Optimization.} Starting from the two initial values $\vartheta^{(0)}_1$ and $\vartheta^{(0)}_2$, maximize $\Psi$ in the unconstrained parameter space $\mathbb{R}^5$ using \texttt{optim} with the \texttt{BFGS} algorithm to obtain $\widehat\vartheta_1$ and $\widehat\vartheta_2$, respectively. Map these back to the original parameters and get $\widetilde\theta_s=\vartheta^{-1}(\widehat\vartheta_s)$ for $s\in\{1,2\}$.
\end{enumerate}

Thus $\widehat f_\pi(x)$ gives us an approximation to the marginal. We would like to sample the $X_i$'s from the $\widehat g_\pi(x_i \mid x_{-i},\widehat\theta_{1:B}) \propto \frac{\prod_{b=1}^B f_\theta(x_i)}{\widehat f_\pi(x)}$. To do that we will be approximating $\widehat g_\pi(x_i \mid x_{-i},\widehat\theta_{1:B})$ using a two component normal distribution which will be used as a proposal in a Metropolis--Hastings algorithm. The rest of this section describes how to obtain this mixture of normal proposal from $\widehat g_\pi(x_i \mid x_{-i},\widehat\theta_{1:B})$. 

Define the negative log-density as $\zeta(x_i):=-\log \widehat g_\pi(x_i \mid x_{-i},\widehat\theta_{1:B})$. We consider a grid 
$\{\xi_j\}_{j=1}^K$ with $K=20$, spanning $[a,b]$ where $a= \min_i X_i, b = \max_i X_i$ and evaluate $\zeta(\xi_j)$ for all $j=1,\dots, 20$. We fit a continuous piecewise–quadratic surrogate of the form
\[
Q(\xi;c,\beta) \;=\; \beta_0
+ \mathbbm{1}\{\xi\le c\}\!\big(\beta_{1L}(\xi-c)+\beta_{2L}(\xi-c)^2\big)
+ \mathbbm{1}\{\xi> c\}\!\big(\beta_{1R}(\xi-c)+\beta_{2R}(\xi-c)^2\big),
\] where $\beta = (\beta_0, \beta_{1L}, \beta_{2L}, \beta_{1R}, \beta_{2R})$, to these $\zeta$ evaluations by minimizing the objective function
\[
\text{SSE}(c) := \min_{\beta}\sum_{j=1}^K\{\zeta(\xi_j)-Q(\xi_j;c,\beta)\}^2.
\]
For each candidate value of $c$, the optimum $\widehat\beta$ can be obtained by a least squares algorithm and the optimum value $c$ (say $\widehat c$) is chosen via a grid search on $400$ different values. This yields coefficients $(\widehat\beta_0,\widehat \beta_{1L},\widehat\beta_{2L},\widehat \beta_{1R},\widehat\beta_{2R})$ at breakpoint $\widehat c$. 

For each arm $s\in\{L,R\}$, we map the quadratic to a Gaussian as follows:
\[
v_s=\max\left\{\epsilon, \frac{1}{2\widehat\beta_{2s}}\right\}, \qquad m_s=\widehat c-\widehat \beta_{1s}v_s\]
with $\epsilon=10^{-8}$. To obtain the mixture weights, we match the magnitude of $Q$ at the two means $m_L$ and $m_R$ with that of a two-component mixture of normal distribution with means $m_L$ and $m_R$ and variances $v_L$ and $v_R$ by solving the following equation for $p$ and $q$: 
\[
\begin{bmatrix}
\phi(m_L;m_L,v_L) & \phi(m_L;m_R,v_R)\\
\phi(m_R;m_L,v_L) & \phi(m_R;m_R,v_R)
\end{bmatrix}
\begin{bmatrix} p\\ q\end{bmatrix}
=
\begin{bmatrix} \exp\left(-Q\left(m_L; \widehat c, \widehat \beta\right)\right)\\ \exp\left(-Q\left(m_R; \widehat c, \widehat \beta\right)\right)\end{bmatrix}.
\]
where $\phi(\,\cdot\,;m,v)$ denotes the pdf of a Gaussian distribution with mean $m$ and variance $v$. Finally, we set the weight (of the $L$ component) as 
$$w = \frac{p}{p+q}.$$
So, the proposal $x_\text{prop}$ is drawn from the two–component Gaussian mixture density
\[
w\,\phi(\cdot;\mu_L,v_L)+(1-w)\,\phi(\cdot;\mu_R,v_R).
\]
This yields the following Metropolis--Hastings algorithm with the permuted serial sampler:
\begin{enumerate}
\item \emph{Initialization.} Draw $m_0\in\{0,\dots,M\}$ uniformly at random, and set
\begin{align*}
&\widetilde X^{(m_0)} \;\leftarrow\; X.
\end{align*}
\item \emph{Iterations.} For $t=m_0+1,\dots,M$, for each \(i=1,\dots,n\) draw $x_{\text{prop}}$ from the density $w\,\phi(\cdot;\mu_L,v_L)+(1-w)\,\phi(\cdot;\mu_R,v_R)$, $u \sim \text{Unif}(0,1)$ and set \[\widetilde X_{i}^{(t)} =
    \begin{cases}
        x_{\text{prop}}, & \text{if } u \leq \alpha, \\
        \widetilde X_{i}^{(t-1)}, & \text{otherwise},
    \end{cases}\] where
\begin{align*}
    \alpha &= \min\!\left\{1,\ 
    \frac{g_{\pi}\!\left(x_{\text{prop}} \mid x_{-i}, \widehat{\theta}_{1:B}\right)}
         {\widehat g_{\pi}\!\left(X_{i}^{(t-1)} \mid x_{-i}, \widehat{\theta}_{1:B}\right)}
    \cdot
    \frac{w\,\phi(X_i^{(t-1)};\mu_L,v_L)+(1-w)\,\phi(X_{i}^{(t-1)};\mu_R,v_R)}
         {w\,\phi(x_{\text{prop}};\mu_L,v_L)+(1-w)\,\phi(x_{\text{prop}};\mu_R,v_R)}
    \right\}
\end{align*}
and $x_{-i} = \bigl(\widetilde X_{<i}^{(t)},\ \widetilde X_{>i}^{(t-1)}\bigr)$.
Similarly, for $t=m_0-1,\dots,0$, for each $i=n,\dots,1$, we draw $x_{\text{prop}}$ from the density $w\,\phi(\cdot;\mu_L,v_L)+(1-w)\,\phi(\cdot;\mu_R,v_R), 
    \ u \sim \text{Unif}(0,1)$ and set $$\widetilde X_{i}^{(t)} =
    \begin{cases}
        x_{\text{prop}}, & \text{if } u \leq \alpha, \\
        \widetilde X_{i}^{(t-1)}, & \text{otherwise},
    \end{cases}$$ as before with the only difference that $x_{-i} = \bigl(\widetilde X_{<i}^{(t+1)},\ \widetilde X_{>i}^{(t)}\bigr).$
\item \emph{Output.} Discard $\widetilde X^{(m_0)}$ and return copies $\widetilde{X}^{(0)},\dots,\widetilde{X}^{(m_0-1)},\widetilde{X}^{(m_0+1)}, \dots, \widetilde{X}^{(M)}$.
\end{enumerate}

\subsection{Rank-1 matrix}\label{app: rank one matrix}
This section gives the implementation details for the rank-1 matrix experiment presented in \Cref{sec: simulation rank one}.

\emph{Sampling from the posterior:}\ 
Recall that our prior is the multivariate standard normal distribution, $U,V\overset{\textnormal{i.i.d.}}{\sim} \mathcal{N}(\vec{0}_n,\mathbb{I}_n)$. 
While the posterior distribution of $(U,V)$---that is, the distribution of $(U,V)\mid X$---is challenging to work with, it is straightforward to compute the conditionals of the posterior distribution: we have
\begin{equation}\label{eqn:U_given_V_X}
U \mid X, V \sim \mathcal{N}\!\left(
\frac{4 X V}{4 \|V\|_2^2 + 1},
\ \frac{1}{4 \|V\|_2^2 + 1}\,\mathbb{I}_n
\right)
\end{equation}
and
\begin{equation}\label{eqn:V_given_U_X}V\mid X,U \sim \mathcal{N}\left( \frac{4 X^\top U}{4 \|U\|^2_2+1}, \frac{1}{4 \|U\|^2_2+1} \mathbb{I}_n\right).\end{equation}
Thus, we can easily sample from the posterior using Gibbs sampling. We begin by initializing $U,V$ via a rank-1 approximation to $X$: writing $u,v\in\mathbb{R}^n$ as the leading left and right singular vectors of $X$, respectively, we initialize with 
\[U = \sqrt{n} \cdot u, \quad V = \sqrt{n}\cdot v,\]
and iterate the Gibbs sampler,
\[\begin{cases}\textnormal{Sample $U\mid V,X$ according to distribution~\eqref{eqn:U_given_V_X}},\\
\textnormal{Sample $V\mid U,X$ according to distribution~\eqref{eqn:V_given_U_X}}.\end{cases}\]
We use a burn-in of $500$ steps, and then extract samples $\widehat\theta_1,\dots,\widehat\theta_B$ at every tenth step.

\emph{Sampling the copies:}\ As in earlier examples, we will need to use a Laplace approximation for the marginal distribution of $X$. However, in this specific setting, we will need to proceed a bit differently---this is because the negative log likelihood of $(U,V)\mid X$ is a non-convex function, and indeed has issues of identifiability, as, e.g., the likelihood takes equal values at $(u,v)$ and $(-u,-v)$. Instead, we will first marginalize over $U$ exactly, and then use a Laplace approximation for marginalizing over $V$. 

First, we calculate the distribution of $X\mid V$, i.e., we marginalize over $U$: using $X_j$ to denote the $j$th row of $X$, we have
\[X_j\mid V \overset{\textnormal{i.i.d.}}{\sim}\mathcal{N}\left(0,0.25\,\mathbb{I}_n+VV^\top\right). \]
The distribution of $X\mid V=v$ therefore has conditional density
\begin{align*}
f_v(x)&= \frac{1}{(2\pi)^{n^2/2}\,\det(0.25\,\mathbb{I}_n+vv^\top)^{n/2}}
   e^{-\frac{1}{2}\sum_{j=1}^n x_j^\top(0.25\,\mathbb{I}_n+vv^\top)^{-1}x_j}\\
&= \frac{1}{(2\pi)^{n^2/2}\,\bigl[(0.25)^n(1+4\|v\|_2^2)\bigr]^{n/2}}
   e^{-\frac{1}{2}\sum_{j=1}^n x_j^\top\Bigl(4\mathbb{I}_n-\frac{16}{1+4\|v\|_2^2}vv^\top\Bigr)x_j}\\
&= \frac{4^{\,n^2/2}}{(2\pi)^{n^2/2}\,(1+4\|v\|_2^2)^{n/2}}
   e^{-2\|x\|_{\mathrm F}^2+\frac{8}{1+4\|v\|_2^2}\,\|x v\|_2^2}.
\end{align*}
where $\|\cdot\|_{\mathrm{F}}$ is the Frobenius norm, and $x_j$ now denotes the $j^{\text{th}}$ row of $x\in\mathbb{R}^{n\times n}$. Writing $X = A\, \text{diag}\{d\}\, B^\top$ as the singular value decomposition of $X$ where $A,B \in \mathbb{R}^{n \times n}$, we can rewrite the marginal as follows:

\begin{align}
\bar{f}_\pi(x) & =\mathbb{E}_{V \sim \mathcal{N}\left(0, \mathbb{I}_n\right)}\left[f_V(x)\right] \notag\\
& =\mathbb{E}_{V \sim \mathcal{N}\left(0, \mathbb{I}_n\right)}\left[\frac{4^{n^2/2}}{(2 \pi)^{n^2 / 2}\left(1+4\|V\|_2^2\right)^{n / 2}} e^{-2\|x\|_{\mathrm{F}}^2+8\frac{\left\|A \cdot \operatorname{diag}\{d\} \cdot B^{\top} V\right\|_2^2}{1+4\|V\|_2^2}}\right] \notag\\
& =\mathbb{E}_{V \sim \mathcal{N}\left(0, \mathbb{I}_n\right)}\left[\frac{4^{n^2/2}}{(2 \pi)^{n^2 / 2}\left(1+4\|V\|_2^2\right)^{n / 2}} e^{-2\|x\|_{\mathrm{F}}^2+8\frac{\left\| \operatorname{diag}\{d\} V\right\|_2^2}{1+4\|V\|_2^2}}\right] \notag\\
&= \mathbb{E}_{W_1,\dots,W_n \stackrel{\text{iid}}{\sim} \chi^2_1}\left[\frac{4^{n^2/2}}{(2 \pi)^{n^2 / 2}\left(1+4 \sum_{i=1}^n W_i\right)^{n / 2}} e^{-2\|x\|_{\mathrm{F}}^2+8\frac{\sum_{i=1}^n W_i d_i^2}{1+4\sum_{i=1}^n W_i}}\right], \label{eq: rank_one_marginal_with_w}
\end{align}
where the second-to-last step holds due to rotational invariance of the standard normal distribution and the $\ell_2$ norm. In the last step, we reparametrize the integrand in~\eqref{eq: rank_one_marginal_with_w} in terms of $W$ instead of $V$ (i.e., $W_i\sim\chi^2_1$ replaces $V_i^2$). This reparametrization addresses the rotational and sign invariances of $f$ in terms of $V$, thereby avoiding complications in the integration arising from multimodality. Before proceeding with the Laplace approximation, we again reparametrize $t_i = \log W_i$; otherwise the $\chi^2_1$ prior for $W_i$ (and hence the unnormalized posterior integrand) is unbounded at $0$, violating the regularity conditions for a Laplace approximation. Since $W_i\sim\chi^2_1$, denoting by $\pi_{W_i}(w_i)$ the $\chi^2_1$ density at $w_i$, the change of variables gives
\[
\pi(t_i)=\pi_{W_i}(e^{t_i})\left|\frac{d}{dt_i}e^{t_i}\right|
=\frac{1}{\sqrt{2\pi}}\,e^{t_i/2}\,e^{-e^{t_i}/2},\qquad t_i\in\mathbb{R},\ i\in \{1,\dots, n\},
\]
hence
\[
\log\pi(t)=\sum_{i=1}^n\left(-\tfrac12\log(2\pi)+\tfrac12 t_i-\tfrac12 e^{t_i}\right),
\quad t=(t_1,\ldots,t_n)^\top .
\]
If we denote
\[
S_1(t):=\sum_{i=1}^n e^{t_i},\qquad S_d(t):=\sum_{i=1}^n d_i^{\,2}e^{t_i},
\]
from~\eqref{eq: rank_one_marginal_with_w}, the conditional density \(f_t(x)\) is
\[
f_t(x)
=\frac{4^{\,n^2/2}}{(2\pi)^{n^2/2}\,\big(1+4S_1(t)\big)^{n/2}}
\exp\!\left(-2\|x\|_F^2+\frac{8\,S_d(t)}{1+4S_1(t)}\right).
\]

Define
\begin{align*}
    \Psi(t;x) &:= \log f_t(x)+\log\pi(t)\\
    &= -\frac{n}{2}\log\!\big(1+4S_1(t)\big)
+\frac{8\,S_d(t)}{1+4S_1(t)}
+\sum_{i=1}^n\!\left(\tfrac{1}{2} t_i-\tfrac{1}{2} e^{t_i}\right)
+\mathrm{const}(x)
\end{align*}
where \(\mathrm{const}(x)=\tfrac{n^2}{2}\log 4-\tfrac{n^2}{2}\log(2\pi)-2\|x\|_F^2\) does not
depend on \(t\).
Introduce the following notations:
\[
c(t)=1+4S_1(t),\qquad N_i(t)=c(t)\,d_i^{\,2}-4S_d(t).
\]
The gradient is then given by
\[
\nabla_t\Psi(t;x)
= \left(\frac{\partial\Psi}{\partial t_1},\dots, \frac{\partial\Psi}{\partial t_n}\right)^\top,
\quad\text{where}\quad
\frac{\partial\Psi}{\partial t_j}
=\frac{1}{2}-\frac{1}{2} e^{t_j}-\frac{2n}{c(t)}e^{t_j}+\frac{8e^{t_j}}{c(t)^2}\big(c(t)\,d_j^{\,2}-4S_d(t)\big).
\]

\noindent The Hessian of the negative log–posterior is $H(t;x):=-\nabla_t^2\Psi(t;x)$ with
\begin{align*}
    H_{jk}(t;x)
&= -\Big[\frac{8n}{c(t)^2}e^{t_j} e^{t_k}
+\frac{32}{c(t)^2}e^{t_j} e^{t_k}\,(d_j^{\,2}-d_k^{\,2})
-\frac{64}{c(t)^3}e^{t_j} e^{t_k}\,N_j(t)\Big], & j\neq k,\\
H_{jj}(t;x)
&=\frac{1}{2} e^{t_j} +2n \left(\frac{e^{t_j}}{c(t)}-\frac{4 e^{2t_j}}{c(t)^2}\right)
-\;8 e^{t_j}\!\left(\frac{N_j(t)}{c(t)^2}-\frac{8 e^{t_j} N_j(t)}{c(t)^3}\right), & j=k.
\end{align*}
We find $\hat t=\arg\max_t \Psi(t;x)$ by using \texttt{optim} in \texttt{R} and denote $\mathbb H :=H(\hat t;x)$.  
The Laplace approximation to the marginal is
\[
\widehat f_\pi(x)
=
(2\pi)^{n/2}\,\det(\mathbb{H})^{-1/2}\,\exp\big\{\Psi(\hat t;x)\big\}.
\]

Using $\widehat{f}_\pi(x)$ we define $\widehat{g}_\pi\left(\;\cdot \mid \widehat{\theta}_{1:B}\right)$ as in \Cref{app:sampling_copies marginal approx} which serves as the target density. Consider the density induced by $\widehat g_\pi(\cdot \mid \widehat\theta_{1:B})$ on each $X_{ij}$, i.e.
\[\widehat g_\pi(x_{ij}\mid x_{-ij},\widehat \theta_{1:B}) \propto \frac{ \prod_{b=1}^Bf_{\widehat\theta_b}(x_{ij})}{\widehat f_\pi(x)^{B-1}}, \]
where $x_{-ij}$ denotes all entries of $x$ except the $(i,j)$-th position. In order to sample each $X_{ij}$ from that density, we use the Metropolis–Hastings algorithm. For the proposal density, we perform a Laplace approximation on this density. Consider $\zeta(x_{ij}) = \log \widehat g_\pi \left(x_{ij}\mid x_{-ij},\widehat\theta_{1:B}\right)$, let
\begin{align*}
    &x_{ij}^\star = \arg\max_{x_{ij}} \zeta(x_{ij}),
\end{align*} and denote the Hessian at $x_{ij}^\star$ by $\zeta^{\prime \prime}(x_{ij}^\star)$. (Note that $x_{ij}^\star$ and $\zeta^{\prime\prime}(x_{ij}^\star)$ are implicitly functions of $x_{-ij}$). The proposal distribution is given by $\mathcal{N}\left(x_{ij}^\star,-\frac{1}{\zeta^{\prime \prime}(x_{ij}^\star)}\right)$. This optimization is performed numerically via \texttt{optim} in \texttt{R} and the resulting Metropolis--Hastings algorithm has average acceptance probability very close to $1$. Putting all of these steps together, we obtain the following permuted serial sampler to sample $\widetilde{X}^{(m)}$'s:
\begin{enumerate}
\item \emph{Initialization.} Draw $m_0\in\{0,\dots,M\}$ uniformly at random, and set
\begin{align*}
&\widetilde X^{(m_0)} \;\leftarrow\; X,
\end{align*}

\item \emph{Iterations.} For $t=m_0+1,\dots,M$, for each \(i=1,\dots,m\) and $j=1,\dots,n$, 

draw $x_{\text{prop}} \sim \mathcal{N}\!\left(x_{ij}^\star,\ - \frac{1}{\zeta^{\prime \prime}(x_{ij}^\star)}\right), 
    \ u \sim \text{Unif}(0,1)$ and set $$\widetilde X_{ij}^{(t)} =
    \begin{cases}
        x_{\text{prop}}, & \text{if } u \leq \alpha, \\
        \widetilde X_{ij}^{(t-1)}, & \text{otherwise},
    \end{cases}$$ where
\begin{align*}
    \alpha &= \min\!\left\{1,\ 
    \frac{\widehat g_{\pi}\!\left(x_{\text{prop}} \mid x_{-ij}, \widehat{\theta}_{1:B}\right)}
         {\widehat g_{\pi}\!\left(X_{ij}^{(t-1)} \mid x_{-ij}, \widehat{\theta}_{1:B}\right)}
    \cdot
    \frac{\phi\!\left((X_{ij}^{(t-1)} - x_{ij}^\star)\sqrt{- \zeta^{\prime \prime}(x_{ij}^\star)}\right)}
         {\phi\!\left((x_{\text{prop}} - x_{ij}^\star)\sqrt{- \zeta^{\prime \prime}(x_{ij}^\star)}\right)}
    \right\}, \\
    x_{-ij} &= \bigl(\widetilde X_{<i,1:n}^{(t)},\ \widetilde X_{i,<j}^{(t)},\ \widetilde X_{i,>j}^{(t-1)},\ \widetilde X_{>i,1:n}^{(t-1)}\bigr).
\end{align*}
Similarly, for $t=m_0-1,\dots,0$, for each $j=n,\dots,1$ and $i=m,\dots,1$, we draw $x_{\text{prop}} \sim \mathcal{N}\!\left(x_{ij}^\star,\ - \frac{1}{\zeta^{\prime \prime}(x_{ij}^\star)}\right), 
    \ u \sim \text{Unif}(0,1)$ and set $$\widetilde X_{ij}^{(t)} =
    \begin{cases}
        x_{\text{prop}}, & \text{if } u \leq \alpha, \\
        \widetilde X_{ij}^{(t-1)}, & \text{otherwise},
    \end{cases}$$ as before with the only difference that $x_{-ij} = \bigl(\widetilde X_{<i,1:n}^{(t)},\ \widetilde X_{i,<j}^{(t)},\ \widetilde X_{i,>j}^{(t+1)},\ \widetilde X_{>i,1:n}^{(t+1)}\bigr).$
\item \emph{Output.} Discard $\widetilde X^{(m_0)}$ and return copies $\widetilde{X}^{(0)},\dots,\widetilde{X}^{(m_0-1)},\widetilde{X}^{(m_0+1)}, \dots, \widetilde{X}^{(M)}$.
\end{enumerate}

\subsection{Group-sparse regression}\label{app: group sparsity}

This section gives the implementation details for the group-sparse regression experiment presented in \Cref{sec: simulation group sparsity}.

\emph{Sampling from the posterior.}\
We restate the priors defined in Section~\ref{sec: simulation group sparsity} here with the general parameters as follows:
\begin{align*}&g^\star \sim \textnormal{Unif(\{1,\dots, G\})},\\
&\beta_{I_{g^\star}}\sim \mathcal{N}(\vec{0}_{|I_{g^\star}|},I_{|I_{g^\star}|}),\\
&\beta_{I_{g}} =\vec{0}_{|I_g|}\ \forall \ g \neq g^\star.\end{align*}
Under this prior, the posterior distribution of $\beta$ can be derived with the following hierarchical structure: defining
\[D_g(X) = |A_g|^{-\frac{1}{2}}\exp\left(\frac{1}{2}b_g^\top  A_g^{-1}b_g - \frac{1}{2}X^\top X\right) \]
for each $g\in[G]$, where 
\[A_g = Z_{I_g}^\top  Z_{I_g} + I_{d_g}, \quad b_g = Z_{I_g}^\top  X, \]
we first sample the active group $g^\star$ as
\[\mathbb{P}(g^\star=g\mid X) \propto D_g(X),\]
then sample $\beta\mid g^\star,X$ as
    \begin{align*}
        &\beta_{I_{g^\star}} \mid g^\star, X\sim \mathcal{N}\left(A_{g^\star}^{-1}b_{g^\star}, A_{g^\star}^{-1}\right),\\
        &\beta_{I_g}\mid g^\star,X = 0 \ \forall \ g\neq g^\star.    \end{align*}

\emph{Sampling the copies:}\ In this case, we can evaluate the marginal of $X$ exactly. Up to normalizing constants,
$$\bar f_\pi(X) = \frac{1}{G}\sum_{g = 1}^G \int_{\mathbb{R}^{d_g}} \exp\left(-\frac{1}{2}||X - Z\beta||^2 - \frac{1}{2}||\beta_{I_g}||^2 \right) \mathsf{d}\beta_{I_g}= \frac{1}{G}\sum_{g=1}^G D_g(X).
$$ In order to sample from the final sampling density $g_\pi(\;\cdot \mid \widehat{\theta}_{1:B})$ as in~\eqref{eqn:conditional_density_X_approx}, we use a Laplace approximation as the proposal in a Metropolis--Hastings sampler. Our goal is to approximate the conditional sampling density of $X_{i}$, and we will update the $X_i$'s one at a time. Since this conditional density is proportional to $g_\pi(X \mid \widehat{\theta}_{1:B})$, we define
$$\zeta(x_i) = -\frac{1}{2} \sum_{b=1}^B (x_i-Z_i^\top\widehat{\beta}_{b})^2 - (B-1)\log\left(\frac{1}{G}\sum_{g=1}^G D_g(x) \right).$$
Taking the derivatives, we can find the maximum and also the Hessian which can then be used to perform a Laplace approximation. 
\begin{align*}
\zeta^\prime(x_i) &= -\sum_{b=1}^B (x_i-Z_i^\top\widehat{\beta}_{b}) - (B-1)\frac{\sum_{g=1}^G \frac{\partial}{\partial x_i} D_g(x)}{\sum_{g=1}^G D_g(x)},\\
\zeta^{\prime\prime}(x_i) &= \frac{B-1}{\left(\sum_{g=1}^G D_g(x)\right)^2} \left[\left(\sum_{g=1}^G D_g(x) \right) \left(\sum_{g=1}^G \frac{\partial ^2}{\partial x_i^2}D_g(x)\right) - \left(\sum_{g=1}^G \frac{\partial}{\partial x_i} D_g(x)\right)^2 \right],
\end{align*}
where $Z_i$ is the $i^\textnormal{th}$ row of the covariate matrix $Z$ and
\begin{align*}
&\frac{\partial}{\partial x_i}D_g(x) = \left[D_g(x) \left(Z_{I_g}A_g^{-1}b_g - x\right)\right]_i,  \\
&\frac{\partial^2}{\partial x_i^2}D_g(x) = \left[D_g(x)\left(\frac{1}{\sigma^2}Z_{I_g}A_g^{-1}Z_{I_g}^\top  -\mathbb{I}_{n}\right)+\left(Z_{I_g} A_g^{-1}b_g - x\right)D_g^{\prime}(x)^\top\right]_{ii}. 
\end{align*}
The optimization is carried out numerically by \texttt{optim} in \texttt{R} to find the maximum $x_i^\star$ and then the approximated Gaussian density $\mathcal{N}\left(x_i^\star, -\frac{1}{\zeta^{\prime\prime}(x_i^\star)}\right)$ is used as the proposal for our Metropolis--Hastings sampling. The acceptance probability in the Metropolis--Hastings stays close to $1$. We use the permuted serial sampler, as follows:
\begin{enumerate}
\item \emph{Initialization.} Draw $m_0\in\{0,\dots,M\}$ uniformly at random, and set
\begin{align*}
&\widetilde X^{(m_0)} \;\leftarrow\; X,
\end{align*}
\item \emph{Iterations.} For $t=m_0+1,\dots,M$, for each \(i=1,\dots,n\) draw $x_{\text{prop}} \sim \mathcal{N}\left(x_{i}^\star,-\frac{1}{\zeta^{\prime \prime}(x_{i}^\star)}\right)$, $u \sim \text{Unif}(0,1)$ and set $$\widetilde X_{i}^{(t)} =
    \begin{cases}
        x_{\text{prop}}, & \text{if } u \leq \alpha, \\
        \widetilde X_{i}^{(t-1)}, & \text{otherwise},
    \end{cases}$$ where
\begin{align*}
    \alpha &= \min\!\left\{1,\ 
    \frac{g_{\pi}\!\left(x_{\text{prop}} \mid x_{-i}, \widehat{\theta}_{1:B}\right)}
         {g_{\pi}\!\left(X_{i}^{(t-1)} \mid x_{-i}, \widehat{\theta}_{1:B}\right)}
    \cdot
    \frac{\phi\!\left((X_{i}^{(t-1)} - x_{i}^\star)\sqrt{-\zeta^{\prime \prime}(x_{i}^\star)}\right)}
         {\phi\!\left((x_{\text{prop}} - x_{i}^\star)\sqrt{-\zeta^{\prime \prime}(x_{i}^\star)}\right)}
    \right\}
\end{align*}
    and $x_{-i} = \bigl(\widetilde X_{<i}^{(t)},\ \widetilde X_{>i}^{(t-1)}\bigr).$
Similarly, for $t=m_0-1,\dots,0$, for each $i=n,\dots,1$, we draw $x_{\text{prop}} \sim \mathcal{N}\!\left(x_{i}^\star,\ -\frac{1}{\zeta^{\prime \prime}(x_{i}^\star)}\right), 
    \ u \sim \text{Unif}(0,1)$ and set $\widetilde X_{i}^{(t)} =
    \begin{cases}
        x_{\text{prop}}, & \text{if } u \leq \alpha, \\
        \widetilde X_{i}^{(t-1)}, & \text{otherwise},
    \end{cases}$ as before with the only difference that $x_{-i} = \bigl(\widetilde X_{<i}^{(t+1)},\ \widetilde X_{>i}^{(t)}\bigr).$
\item \emph{Output.} Discard $\widetilde X^{(m_0)}$ and return copies $\widetilde{X}^{(0)},\dots,\widetilde{X}^{(m_0-1)},\widetilde{X}^{(m_0+1)}, \dots, \widetilde{X}^{(M)}$.
\end{enumerate}

\subsection{Linear spline regression}\label{app: linear spline}

This section gives the implementation details for the linear spline regression experiment presented in Section~\ref{sec: simulation linear spline}.

\emph{Sampling from the posterior:}\ Following~\eqref{eq: spline_reparametrize}, with $k=1$ (i.e., one knot) the linear spline model can be represented as 
$$X = \gamma_0 + \gamma_1 Z +  \gamma_{2} b_1(Z) + \epsilon$$
which can be further written as: 
$$X = h_t(Z) \gamma + \epsilon,$$ where $$h_t(Z) = \left(\vec{1}_n, Z, b_1(Z)\right)\in\mathbb{R}^{n\times 3} \quad\textnormal{ and }\quad\gamma = (\gamma_0, \gamma_1, \gamma_2 )^\top.$$
Let $Z_{(i)}$ denote the $i$-th order statistic of $Z_1,\dots,Z_n$ and for notational convenience we shall denote $Z_{(0)} = -\infty$ and $Z_{(n+1)} = \infty$. We shall use $X_{(i)}$ to denote the response corresponding to covariate $Z_{(i)}$. Recall that we are using the prior distribution
\begin{align*}
&\gamma_0,\gamma_1,\gamma_2\overset{\textnormal{i.i.d.}}{\sim} \mathcal{N}(0,1),\\
&t_1\sim \mathcal{N}(0,1).
\end{align*}
To draw the posterior samples, we use a Gibbs sampling algorithm where the conditional distributions are as follows. First, the conditional distribution of $\gamma$ is given by
\begin{align}\label{eq: spline_post_beta}
    (\gamma \mid t, Z, X) &\sim \mathcal{N}\left(\mu_\gamma, V_\gamma\right),  \\
    \textnormal{where }V_\gamma = \left(4 h_t(Z)^\top h_t(Z) + I_3\right)^{-1} &\quad \textnormal{and} \quad \mu_\gamma = V_\gamma \left(4 h_t(Z)^\top X\right). \notag
\end{align}
Next, the conditional distribution of $t_1$ has density
\begin{align*}
    P(t_1 \mid \gamma, X, Z) &\propto \phi\left(\frac{\|X - h_t(Z)\gamma\|}{\sigma}\right)\phi\left(\frac{t_1}{\tau_2}\right)\\
    &\propto \exp\left(-2 \sum_{i=1}^{n} \left(X_i - \gamma_{0} - \gamma_{1} Z_i - \gamma_{2} b_1(Z_i)\right)^2 \right)\exp\left(-\frac{t_1^2}{2}\right).
\end{align*}
Recalling that $b_1(z) = (z-t_1)_+$, we note that in this distribution, for any $i$, when $t_1 \in (Z_{(i)},Z_{(i+1)})$, the density $P(t_1 \mid \gamma, X, Z)$ is 
\[\propto \exp\left(-2 \sum_{i'=i+1}^{n} \left(X_{(i')} - \gamma_{0} - \gamma_{1} Z_{(i')} - \gamma_{2} (Z_{(i')} - t_1)\right)^2 \right)\exp\left(-\frac{t_1^2}{2}\right),\]
and is therefore proportional to the following Gaussian distribution:
\begin{align}\label{eq: interval of t_j}
    P(t_1 \mid t_1 \in (Z_{(i)}, Z_{(i+1)}), \gamma, X, Z) \propto \mathcal{N}(\mu_{t,i}, \sigma_{t,i}^2),
\end{align}
\begin{align*}
\text{where, } && \mu_{t,i} &= \left(4(n-i)\, \gamma_{2}^2 + 1\right)^{-1}
\left(4\gamma_{2} \sum_{i'=i+1}^n\left(  X_{(i')} - \gamma_{0} - \gamma_{1} Z_{(i')} -\gamma_{2}\,Z_{(i')}\right) \right),\\
&& \sigma_{t,i}^2 &= \left(4(n-i)\,\gamma_{2}^2 + 1\right)^{-1}.
\end{align*}
The probability that $t_1\in (Z_{(i)},Z_{(i+1)})$ can be obtained by evaluating the following integral
\begin{align}\label{eq: t_j given interval}
    \mathbb{P}(t_1 &\in (Z_{(i)}, Z_{(i+1)})\mid \gamma, X, Z)\notag \\
    &= \int_{t_1 = Z_{(i)}}^{Z_{(i+1)}} e^{-2\sum_{i^\prime = 1}^{i} \left(X_{(i')} - \gamma_{0} - \gamma_{1} Z_{(i')}\right)^2 -2\sum_{i^\prime = i+1}^{n}\left(X_{(i')} - \gamma_{0} - \gamma_{1} Z_{(i')}-\gamma_{2}(Z_{(i^\prime)}-t_1)\right)^2
    -\frac{1}{2}t_1^2} \, \mathsf{d}t_1 \notag\\
    &=: w_i.
\end{align}
This integration can be solved by expressing the exponent in the integrand as a quadratic in $t_1$ and then using the Gaussian CDF. This enables us to sample from the posterior of $t_1$ by first sampling the interval in which $t_1$ lies using the above probability weights and subsequently drawing $t_1$ from a truncated normal distribution.
We will represent this sampling distribution concisely as follows:
\begin{align}\label{eq: complete sampling distribution t_j}
    t_1\mid  Z, X, \gamma \sim \sum_{i=0}^{n} w_i\, \mathrm{TN}(\mu_{t,i} , \sigma_{t,i}^2,Z_{(i)},Z_{(i+1)}),
\end{align}
where TN$(\mu,\sigma^2,a,b)$ is the truncated normal distribution---that is, the distribution $\mathcal{N}(\mu,\sigma^2)$ truncated to the interval $[a,b]$.
This results in the following Gibbs sampling algorithm to sample from the posterior distribution of the parameters. We use the \texttt{R} package \texttt{segmented}~\citep{segmented_package} to find the position of the knots and initialize $t^{(0)}$ at that point. We initialize $\gamma^{(0)}$ as the coefficient vector from the regression of $X$ on $h_{t^{(0)}}$. For $b=1,\dots, B$,
\begin{enumerate}
    \item Generate $\gamma^{(b)}$ from $\gamma \mid t_1^{(b-1)},Z,X$ as in \Cref{eq: spline_post_beta}.
    \item Generate $t^{(b)}_1$ from $t_1\mid Z, X, \gamma^{(b)} $ as in \Cref{eq: complete sampling distribution t_j}.
\end{enumerate}
After the burn-in of \(B_0=500\), we extract $B=25$ posterior samples \(\{t^{(b)},\gamma^{(b)}\}\) at every tenth step.

\emph{Sampling the copies:}\ Our first step is to approximate the marginal $\bar{f}_\pi(x)$. In this case, we first marginalize out $\gamma$ from the distribution of $X\mid t, \gamma$ as follows:
\begin{align*}
    & X \mid t_1, \gamma \sim \mathcal{N}\left(h_t(Z) \gamma, 0.25 \mathbb{I}_n\right)\\
    \implies & X \mid t_1 \sim \mathcal{N}\left(\vec{0}_n, h_t(Z) h_t(Z)^\top  + 0.25 \mathbb{I}_n\right).
\end{align*}

Now consider the joint density of $(X,t_1)$. Since our prior on $t_1$ is $\mathcal{N}(0,1)$, the joint density is given by
\[\Psi(x,t_1) = \frac{1}{\sqrt{(2\pi)^n |h_t(Z) h_t(Z)^\top  + 0.25 \mathbb{I}_n|}} e^{-\frac{1}{2}x^\top(h_t(Z) h_t(Z)^\top  + 0.25 \mathbb{I}_n)^{-1} x} \cdot \frac{1}{\sqrt{2\pi}}e^{-t_1^2/2}\] and the marginal is then given by $\bar{f}_\pi(x) = \int \Psi(x,t_1)\;\mathsf{d}t_1$. Note that evaluating $\Psi(x,t_1)$ at a single value $t_1$ is simple, but integrating this quantity is challenging. We will therefore take an approximation: first re-define $Z_{(0)} = -C$ and $Z_{(n+1)} = C$ (for a large value $C$ chosen so that the tail mass of $\Psi$ outside $[-C,C]$ is negligible), and let $Z_{(1)},\dots,Z_{(n)}$ be as before. We then define a grid $z_0\leq \dots \leq z_{K(n+1)}$ where $z_{Ki}=Z_{(i)}$ for each $i$, and the points $z_{K(i-1)},\dots,z_{Ki}$ form an equally spaced grid from $Z_{(i-1)}$ to $Z_{(i)}$ for each $i$. After evaluating $\Psi(z_0),\dots,\Psi(z_{K(n+1)})$, we approximate the integral with the trapezoid rule:
\[\widehat{f}_\pi(x) = \sum_{i=1}^{n+1}\sum_{k=1}^K \frac{\Psi(x,z_{K(i-1)+k-1}) + \Psi(x,z_{K(i-1)}+k)}{2} \cdot \frac{Z_{(i)}-Z_{(i-1)}}{K}.\]
For our implementation we choose $C=10$ and $K=20$.

Using this we can define the approximate target density $\widehat g_\pi\left(\;\cdot \mid \widehat{\theta}_{1:B}\right)$ as in Section~\ref{app:sampling_copies marginal approx} which serves as the target density. In order to sample from that density, we use the Metropolis–Hastings algorithm. Consider $\zeta(x_{i}) = \log \widehat g_\pi \left(x_{i}\mid x_{-i},\widehat\theta_{1:B}\right)$,
\begin{align*}
    &x_{i}^\star = \arg\max_{x_{i}} \zeta(x_{i}),
\end{align*} and denote the Hessian at $x_{i}^\star$ by $\zeta^{\prime \prime}(x_{i}^\star)$. The proposal distribution is given by $\mathcal{N}\left(x_{i}^\star,-\frac{1}{\zeta^{\prime \prime}(x_{i}^\star)}\right)$. This optimization is performed numerically via \texttt{optim} in \texttt{R}. We use the permuted serial sampler, as follows:

\begin{enumerate}
\item \emph{Initialization.} Draw $m_0\in\{0,\dots,M\}$ uniformly at random, and set
\begin{align*}
&\widetilde X^{(m_0)} \;\leftarrow\; X.
\end{align*}
\item \emph{Iterations.} For $t=m_0+1,\dots,M$, for each \(i=1,\dots,n\) draw $x_{\text{prop}} \sim \mathcal{N}\left(x_{i}^\star,-\frac{1}{\Psi^{\prime \prime}(x_{i}^\star)}\right)$, $u \sim \text{Unif}(0,1)$ and set $$\widetilde X_{i}^{(t)} =
    \begin{cases}
        x_{\text{prop}}, & \text{if } u \leq \alpha, \\
        \widetilde X_{i}^{(t-1)}, & \text{otherwise},
    \end{cases}$$ where
\begin{align*}
    \alpha &= \min\!\left\{1,\ 
    \frac{\widehat g_{\pi}\!\left(x_{\text{prop}} \mid x_{-i}, \widehat{\theta}_{1:B}\right)}
         {\widehat g_{\pi}\!\left(X_{i}^{(t-1)} \mid x_{-i}, \widehat{\theta}_{1:B}\right)}
    \cdot
    \frac{\phi\!\left((X_{i}^{(t-1)} - x_{i}^\star)\sqrt{-\zeta^{\prime \prime}(x_{i}^\star)}\right)}
         {\phi\!\left((x_{\text{prop}} - x_{i}^\star)\sqrt{-\zeta^{\prime \prime}(x_{i}^\star)}\right)}
    \right\}
\end{align*}
    and $x_{-i} = \bigl(\widetilde X_{<i}^{(t)},\ \widetilde X_{>i}^{(t-1)}\bigr)$.
Similarly, for $t=m_0-1,\dots,0$, for each $i=n,\dots,1$, we draw $x_{\text{prop}} \sim \mathcal{N}\!\left(x_{i}^\star,\ -\frac{1}{\zeta^{\prime \prime}(x_{i}^\star)}\right), 
    \ u \sim \text{Unif}(0,1)$ and set $\widetilde X_{i}^{(t)} =
    \begin{cases}
        x_{\text{prop}}, & \text{if } u \leq \alpha, \\
        \widetilde X_{i}^{(t-1)}, & \text{otherwise},
    \end{cases}$ as before with the only difference that $x_{-i} = \bigl(\widetilde X_{<i}^{(t+1)},\ \widetilde X_{>i}^{(t)}\bigr).$
\item \emph{Output.} Discard $\widetilde X^{(m_0)}$ and return copies $\widetilde{X}^{(0)},\dots,\widetilde{X}^{(m_0-1)},\widetilde{X}^{(m_0+1)}, \dots, \widetilde{X}^{(M)}$.
\end{enumerate}

\section{Sensitivity analysis}\label{app: sensitivity analysis}
\subsection{Sensitivity with respect to concentration of prior}\label{app: sensitivity wrt prior}
In this section, we investigate the impact of the concentration of the prior on the performance of aCSS-B.
While aCSS-B maintains validity and matches the oracle's power across our simulations under standard specifications, it is instructive to examine the performance under a wide range of priors; two extreme cases being a highly concentrated prior centred far from the true parameter, and conversely, an extremely diffuse prior. Notably, highly concentrated priors are seldom used in practice without substantial prior information; furthermore, such priors are generally inadvisable for use with aCSS-B.

To get some intuition about the behaviour of our method, recall that the type-I error inflation bound in Theorem \ref{thm: validity} is given by the term $\epsilon(\pi_0)+\frac{\Delta(\pi_0)}{2\sqrt{B}}$. While the first term $\epsilon(\pi_0)$ does not directly depend on the chosen prior $\pi$, $\Delta(\pi_0)$, which is the expected distance between the posteriors $\pi_0(\cdot\mid X)$ and $\pi(\cdot\mid X)$, is substantially impacted by this choice. Since $\pi_0$ is highly concentrated around $\theta_0$, if $\pi$ concentrates around some other $\theta_1$ far away from $\theta_0$, the posterior distribution $\pi(\cdot\mid X)$ might be substantially different from $\pi_0(\cdot\mid X)$ i.e. $\textnormal{d}_{\chi^2}(\pi_0(\cdot\mid X),\pi(\cdot\mid X))$ will be large making $\Delta \gg0$. This would require using a very large $B$ to ensure that $\frac{\Delta(\pi_0)}{2\sqrt{B}}\approx 0$. However, if $B$ is too large, this will end up conditioning on too much information in the distribution of $\widetilde{X}$ which would cause this distribution to be highly concentrated around $X$ and result in low power. Conversely, when $\pi$ is flat or weakly informative, with a sufficient sample size $n$ both the posteriors $\pi_0(\cdot\mid X)$ and $\pi(\cdot\mid X)$ are concentrated around the data-generating parameter $\theta_0$, making $\Delta(\pi_0)$ small. This allows a much smaller $B$ which results in a more powerful test.

To empirically investigate how the choice of prior affects the power and validity of our method, we choose two settings from the ones presented in the main paper---logistic regression (Section~\ref{sec: simulation logistic regression}) and group-sparse regression (Section~\ref{sec: simulation group sparsity})---applying our method with five different prior-parameter choices that control the prior’s flatness.

For logistic regression, we vary the prior scale by considering five values of $\tau$, the standard deviation in $\theta_j \sim \mathcal{N}(0,\tau^2)$: $\tau \in \{0.01, 0.1, 1, 10, 100\}$, with data generated under $\theta_0 = 0.2\,\mathbbm{1}_d$. In practice, highly concentrated priors such as $\tau = 0.1$ or $0.01$ are unlikely to be chosen by practitioners; we include them primarily to assess the behaviour of our method under extreme prior specifications. Across this range---from very concentrated to highly diffuse priors---the resulting aCSS-B procedures remain valid and achieve essentially unchanged power, closely tracking the oracle.

\begin{figure}[ht]
    \centering
    \includegraphics[width=0.6\linewidth]{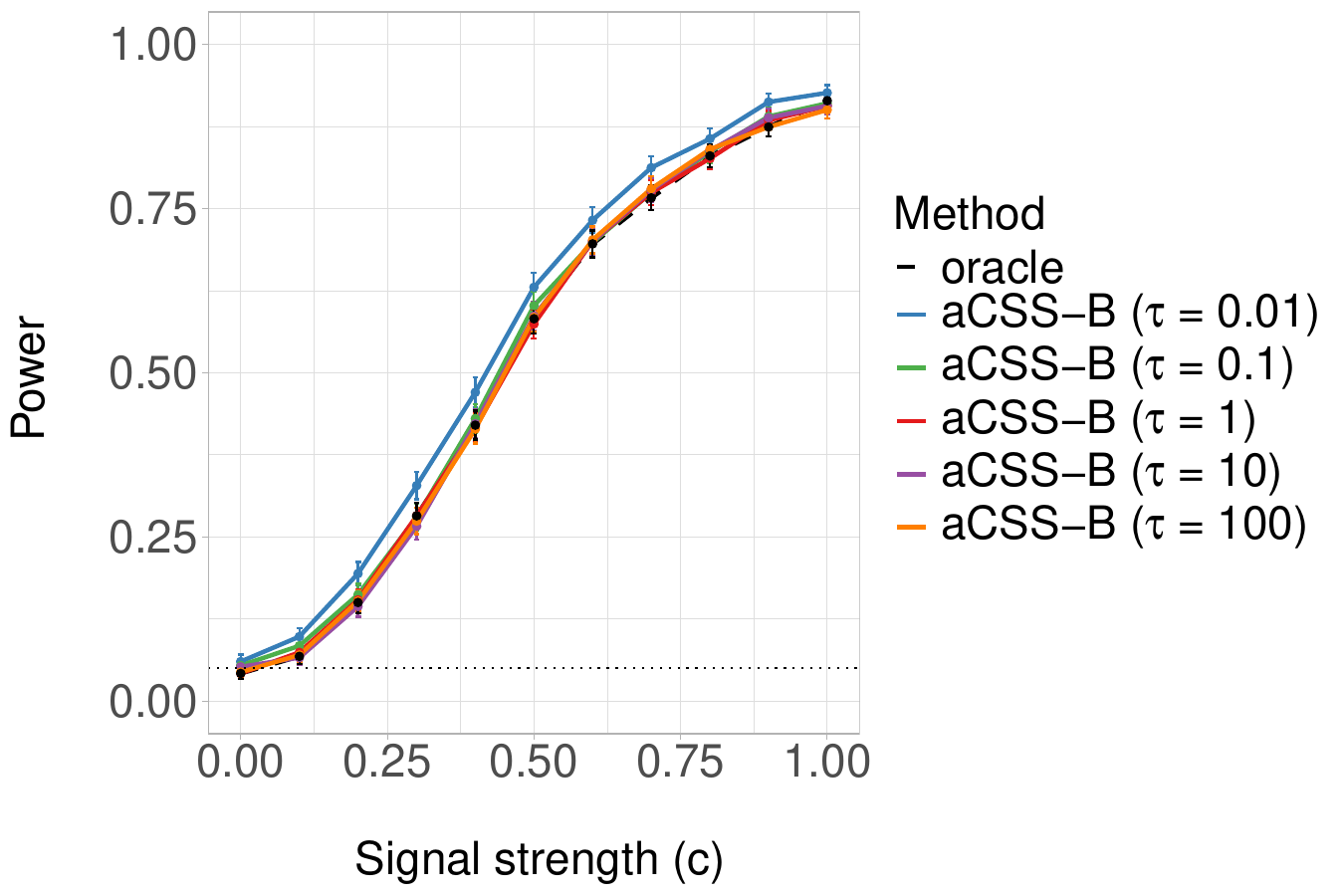}
    \caption{Power of aCSS-B under different prior standard deviations in the logistic regression model of Section~4.1.}
    \label{fig:sensitivity_logistic}
\end{figure}

For group-sparse regression, recall the prior
\begin{align}
    &g^\star \sim \textnormal{Unif}(\{1,\dots, G\}), \nonumber\\
    &\beta_{I_{g^\star}}\sim \mathcal{N}(5\cdot \vec{1}_{|I_{g^\star}|},\tau^2I_{|I_{g^\star}|}),\nonumber\\
    &\beta_{I_{g}} =\vec{0}_{|I_g|}\ \forall \ g \neq g^\star.\label{eq:prior_group_sparsity_sup}
\end{align}
We vary the prior standard deviation over $\tau \in \{0.01, 0.1, 1, 10, 100\}$ and use a fixed null parameter $\beta$ with
\begin{equation}
    \beta_{I_{g_1}} = \vec{1}_{|I_{g_1}|}, \ 
    \beta_{I_{g_2}} = c\,\vec{1}_{|I_{g_2}|}, \ \ 
    \beta_j = 0 \ \forall \ j \in [d]\backslash(I_{g_1} \cup I_{g_2}).\label{eq:coeff_group_sparsity_sup}
\end{equation}
which ensures that the true parameter is systematically separated from the prior mean $5\cdot \vec{1}_{|I_{g_1}|}$---especially for concentrated priors. As in the logistic regression experiment, $\tau=0.01$ and $0.1$ correspond to highly concentrated and practically implausible prior choices.

\begin{figure}[ht]
    \centering
    \includegraphics[width=0.6\linewidth]{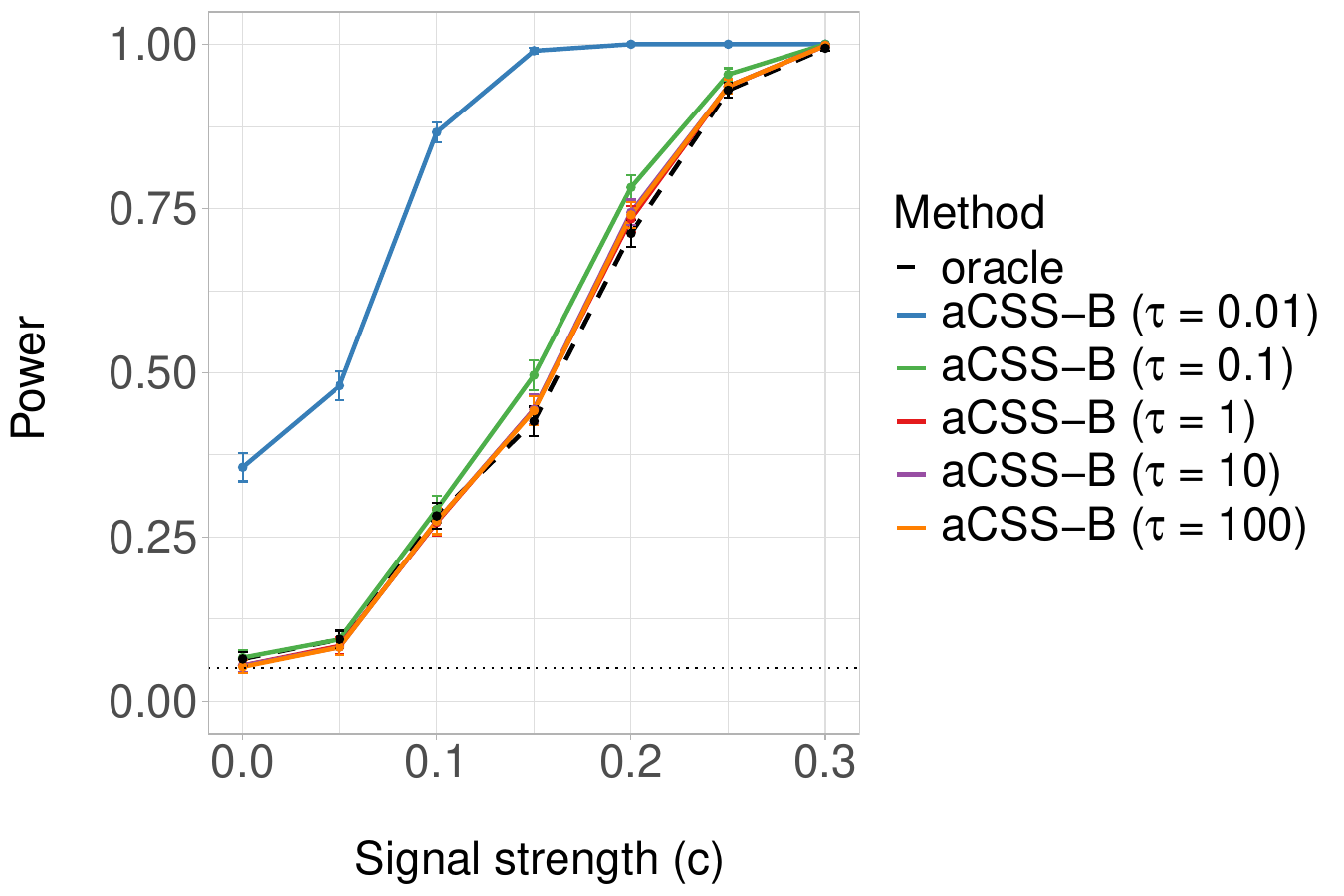}
    \caption{Power of aCSS-B under different prior standard deviations for the active group in the group-sparse model of Section 4.4, with $B=25$ posterior draws. Error bars show $\pm 1$ standard error.}
    \label{fig:sensitivity_group_sparsity_B=25}
\end{figure}

Figure~\ref{fig:sensitivity_group_sparsity_B=25} shows that aCSS-B remains valid for $\tau \in \{0.1, 1, 10, 100\}$ and achieves power comparable to the oracle, but becomes invalid under the extremely concentrated prior $\tau=0.01$.

\subsection{Sensitivity with respect to number of posterior samples}\label{app: sensitivity wrt B}
Section~\ref{app: sensitivity wrt prior} shows that in the group sparse regression example, when using $B=25$ posterior samples, a highly concentrated (around the wrong parameter value) prior such as one with $\tau=0.01$ results in substantial type-I error inflation by the aCSS-B method. Theorem~\ref{thm: validity}  suggests that this should be  mitigated when using a higher number of posterior samples, however, possibly at the cost of power. In this section, we empirically verify this phenomenon in more detail by considering two different values of the prior standard deviation $\tau = \{0.01,1\}$ in the group sparse regression model. These two prior parameters respectively represent highly informative and reasonable choices of prior. The prior mean is $5$ which is quite far away from the true data generating parameter $1$. The data generating distribution and the prior distribution have been specified in detail in~\eqref{eq:prior_group_sparsity_sup} and \eqref{eq:coeff_group_sparsity_sup} respectively and we vary $B$ in the range $\{25,100,250,500,1000\}$.

\begin{figure}[ht]
    \centering
    \includegraphics[width=0.95\linewidth]{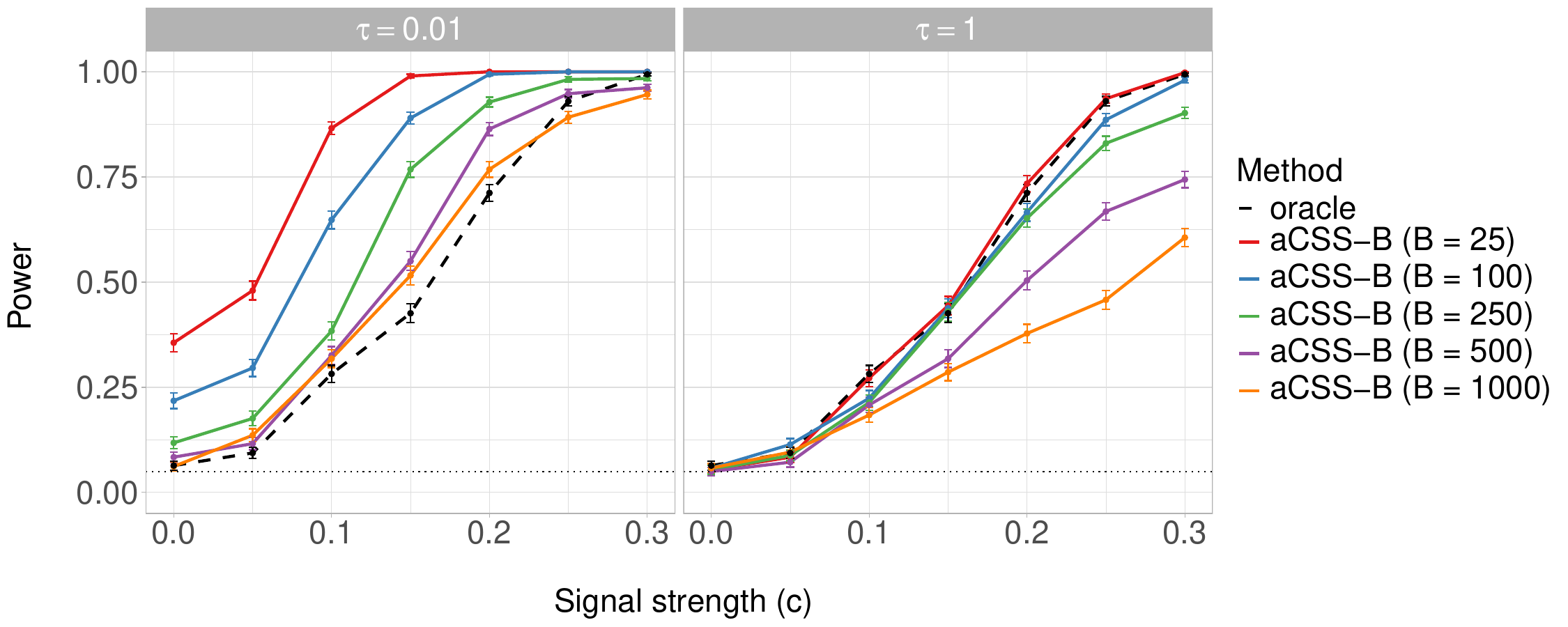}
    \caption{Power of aCSS-B under different choices of $B$ for two different prior standard deviations ($\tau = 0.01$ and $1$) in the group-sparse model of Section 4.4. Error bars show $\pm 1$ standard error.}
    \label{fig:sensitivity_group_sparsity_vs_B}
\end{figure}

Figure~\ref{fig:sensitivity_group_sparsity_vs_B} illustrates that when the prior is overly concentrated far from the true data-generating parameter ($\tau = 0.01$), aCSS-B loses validity for most values of $B$. In this adversarial setting, $B$ must be as large as $1000$ to achieve validity comparable to the oracle test. Conversely, with a more reasonable prior ($\tau = 1$), $B=25$ is sufficient to ensure validity. While increasing $B$ beyond this point does not affect validity, as expected, an extremely large value of $B$ does result in a reduction of statistical power.

\end{document}